\documentclass[aps,prc,preprint,groupedaddress,preprintnumbers,12pt]{revtex4-1}
\usepackage{natbib}
\usepackage{amsmath}
\usepackage{graphicx}
\usepackage{hyperref}
\usepackage{bm}
\allowdisplaybreaks[1]
\newcommand{\heedp}{$^3$He($e,e^\prime p$)$^2$H}
\newcommand{\heedpfour}{$^4$He($e,e^\prime p$)$^3$H}
\newcommand{\heedpfourpol}{$^4$He($\vec{e},e^\prime \vec{p}\,$)$^3$H}
\begin{document}


\title{The $^3$He$(e,e^\prime p)^2$H and $^4$He$(e,e^\prime p)^3$H reactions at high momentum transfer}

\author{W.P.\ Ford$^{\rm a}$}
\author{R.\ Schiavilla$^{\rm b,c}$}
\author{J.W.\ Van Orden$^{\rm b,c}$}
\affiliation{
$^{\rm a}$\mbox{Department of Physics, University of Southern Mississippi, Hattiesburg, MS 39406 }\\
$^{\rm b}$\mbox{Department of Physics, Old Dominion University, Norfolk, VA 23529, USA}\\
$^{\rm c}$\mbox{Jefferson Lab, Newport News, VA 23606, USA}\\
}
\date{\today}
\begin{abstract}
We present updated calculations for observables in the processes \heedp,
\heedpfour, and \heedpfourpol.  This update entails the implementation of
improved nucleon-nucleon ($NN$) amplitudes to describe final state interactions
(FSI) within a Glauber approximation and includes full spin-isospin dependence
in the profile operator.  In addition, an optical potential, which has also been
updated since previous work, is utilized to treat FSI for the \heedpfour ~and 
\heedpfourpol ~reactions.  The calculations are compared with experimental
data and show good agreement between theory and experiment.  Comparisons
are made between the various approximations in the Glauber treatment, including
model dependence due to the $NN$ scattering amplitudes, rescattering contributions,
and spin dependence.   We also analyze the validity of the Glauber approximation at
the kinematics the data is available, by comparing to the results obtained with the
optical potential.
\end{abstract}
\maketitle

\section{Introduction}
\label{sec:intro}
Recent experiments at Jefferson Lab (JLab) have measured cross sections and
polarization observables for the $^3$He($e,e^\prime p$)$^2$H,
$^4$He($e,e^\prime p$)$^3$H, and $^4$He($\vec{e},e^\prime \vec{p}\,$)$^3$H
reactions at intermediate and large momentum
transfers~\cite{Rvachev05,Reitz04,Strauch03,Paolone10,Malace:2011} .
These data have generated considerable interest in the nuclear few-body community,
as attested by the series of papers dealing with the description of the proton-knockout
mechanism and the treatment of final state interactions (FSI) at GeV energies, which
have appeared in the literature in last few years
~\cite{ciofihiko_2,ciofihiko_3,ciofihiko_4,Ciofi2010,Laget2005,Laget200549,Sargsian2005,Sargsian2005_2}.

In the present work, we report on a calculation of the two-body electrodisintegration
cross sections of $^3$He and $^4$He in the wide range of momentum transfers covered
by the JLab experiments.  This study updates, improves, and extends that of
Refs.~\cite{Schiavilla05a,Schiavilla05b}.  As in the earlier work, the nuclear
bound states are represented by non-relativistic wave functions, obtained from realistic
two- and three-nucleon potentials (the Argonne $v_{18}$ two-nucleon~\cite{Wiringa95}
and Urbana IX three-nucleon~\cite{Pudliner95} potentials---the AV18/UIX Hamiltonian)
and FSI between the outgoing proton and recoiling bound cluster are treated either
in the Glauber approximation for the $A$=3 and 4 reactions---with inclusion in the
associated profile operator of the full spin-isospin dependence of the nucleon-nucleon
($NN$) elastic scattering amplitude---or, in the case of the $A$=4 reactions, with an
optical potential.  Important differences between the present work and that
of Refs.~\cite{Schiavilla05a,Schiavilla05b} are that: i) the $NN$ amplitudes are obtained
from the Scattering Analysis Interactive Dial-in (SAID) analysis~\cite{Arndt00,Arndt07,SAIDdata} 
of $pn$ ($pp$) scattering data at lab kinetic energies
ranging from 0.05 (0.05) GeV to 1.3 (3.0) GeV rather than from a parametrization of these
amplitudes valid at forward scattering (at small momentum transfers)~\cite{Wallace81},
and ii) the parameters in the optical potential have been adjusted to reproduce,
in addition to $^3$H($p,p$)$^3$H elastic and $^3$H($p,n$)$^3$He charge-exchange
cross section data, also the induced polarization data recently measured for the
$^4$He($\vec{e},e^\prime \vec{p}\,$)$^3$H reaction at JLab~\cite{Malace:2011}.

This paper is organized as follows.  In Sec.~\ref{sec:fsi}, we briefly discuss our treatment
of FSI both in the Glauber and optical-model approximations, relegating details on
the construction of the Glauber profile operator from the $NN$ SAID amplitudes
to Appendices~\ref{app:NNSAIDtoWallace} and~\ref{app:flab}.  In Sec.~\ref{sec:calc}
we review the bound-state wave functions, the model for the electromagnetic current
operator, and the Monte Carlo methods used in the numerical evaluation of the
relevant matrix elements---these methods have already been described in considerable
detail in Ref.~\cite{Schiavilla05a}.  
In Sec~\ref{sec:observables} we list explicit expressions for the observables of interest to this work.  
Finally, in Sec.~\ref{sec:res} we present
a detailed discussion of the results, including a comparison between
the Glauber and optical-model treatment of FSI in kinematical
regimes where both approaches are expected to be valid, and in Sec.~\ref{sec:concl}
we summarize our conclusions.
\section{Final state interactions}
\label{sec:fsi}

Two different approximations are adopted in the present work to describe 
FSI in the two-body electrodisintegrations
of $^3$He and $^4$He: one is based on the Glauber approach~\cite{Glauber59},
while the other, whose application is limited only to processes involving
$^4$He, relies on an optical potential.  Both approximations have been
discussed in considerable detail in Refs.~\cite{Schiavilla05a}
and~\cite{Schiavilla05b}: each has limitations as to the energy range
where it is expected to be reliable.  For completeness, in this section
we briefly review them, emphasizing those aspects of the approach which have been improved since the
study of Refs.~\cite{Schiavilla05a,Schiavilla05b}.
\subsection{Glauber approach}
\label{sec:glb}

In this approach the wave function of the final $p$+($A-1$) system is
written as
\begin{equation}
\psi(p+^{(A-1)}\!f;{\rm GLB})=
\frac{1}{\sqrt{A}} \sum_{\cal P} \epsilon_{\cal P}\, G(A;1 \dots A-1)\,
{\rm e}^{{\rm i} {\bf p} \cdot {\bf r}_A} \chi_\sigma(A;p)\,
       {\rm e}^{{\rm i} {\bf p}_f \cdot {\bf R}_{1\dots A-1}}
\phi_{\sigma_f}(1\dots A-1;f) \ ,
\label{eq:glb}
\end{equation}
where $\chi_\sigma(p)$ represents a proton in spin state $\sigma$,
$\phi_{\sigma_f}(f)$ denotes the wave function of the ($A-1$)-system
with spin projection $\sigma_f$, and ${\bf R}_{1 \dots A-1}$ is the
center-of-mass position vector of the $A-1$ nucleons in this cluster.
The sum over permutations $\cal P$ of parity $\epsilon_{\cal P}$ ensures
the overall antisymmetry of $\psi(p+^{(A-1)}\!f;{\rm GLB})$.

The operator $G(A;1 \dots A-1)$ inducing FSI can be derived from an
analysis of the multiple scattering series by requiring that the struck
(fast) nucleon (nucleon $A$) is undeflected by rescattering processes,
and that the nucleons in the residual system (nucleons $1, \dots, A-1$)
act as fixed scattering centers~\cite{Wallace81}.  It is expanded as
\begin{equation}
G=1 + \sum_{n=1}^{A-1} (-)^n G^{(n)} \ ,
\end{equation}
where $G^{(n)}$ represents the $n^{\rm th}$ rescattering term, and therefore
for an $A$-body system up to $A-1$ rescattering terms are generally present.
The leading single-rescattering term reads
\begin{equation}
G^{(1)}(A;1\dots A-1)=\sum_{i=1}^{A-1} \theta(z_{iA})
\, \Gamma_{iA}({\bf b}_{iA};s_{iA}) \ ,
\label{eq:glb1}
\end{equation}
where $z_{iA}$ and ${\bf b}_{iA}$ denote the longitudinal and
transverse components of ${\bf r}_i-{\bf r}_A$ relative to
$\hat {\bf p}$, the direction of the nucleon momentum,
\begin{equation}
z_{iA} \equiv {\hat{\bf p}}\cdot ({\bf r}_i-{\bf r}_A) \ , \qquad
{\bf r}_i-{\bf r}_A \equiv {\bf b}_{iA} + z_{iA} \,{\hat{\bf p}} \ ,
\end{equation}
and the step-function $\theta(x)$, $\theta(x)=1$ if $x > 0$ and $\theta(x)=0$ if $x < 0$, prevents
the occurrence of backward scattering for the struck nucleon.  The
\lq\lq profile operator\rq\rq $\ \Gamma_{iA}$, derived from the
$N$$N$ elastic scattering amplitude at the invariant energy
$\sqrt{s_{iA}}$, is discussed below.
The double- and triple-rescattering terms, relevant for the present study
of the $^3$He($e,e^\prime p$)$d$ and $^4$He($e,e^\prime p$)$t$ reactions,
are given by
\begin{equation}
G^{(2)}(A;1\dots A-1)=\sum_{i\ne j=1}^{A-1}
\theta(z_{ij})\, \theta(z_{jA}) \,
 \Gamma_{iA}({\bf b}_{iA};s_{iA})\,
\Gamma_{jA}({\bf b}_{jA};s_{jA}) \ , 
\end{equation}
\begin{equation}
G^{(3)}(A;1\dots A-1)=\sum_{i\ne j\ne k=1}^{A-1}
\theta(z_{ij})\,\theta(z_{jk})\, \theta(z_{kA}) \,
\Gamma_{iA}({\bf b}_{iA};s_{iA})\,
\Gamma_{jA}({\bf b}_{jA};s_{jA})\,
\Gamma_{kA}({\bf b}_{kA};s_{kA}) \ ,
\end{equation}
where the product of $\theta$-functions ensures the correct sequence
of rescattering processes in the forward hemisphere.

The profile operator $\Gamma_{ij}$ is related to the $N$$N$
scattering amplitude, denoted as $F_{ij}({\bf k};s)$, via the Fourier
transform

\begin{equation}
\Gamma_{ij}({\bf b};s) = \frac{1}{ 2 \pi {\rm i}\, p }
\int {\rm d}^2{\bf k}\, {\rm e}^{-{\rm i} {\bf k}\cdot {\bf b}}
F_{ij}({\bf k};s) \ ,
\label{eq:prof}
\end{equation}
where, in the eikonal limit, the momentum transfer ${\bf k}$ is
perpendicular to ${\bf p}$.  The isospin symmetry of the strong
interactions allows one to express $F_{ij}$ as

\begin{equation}
F_{ij}=F_{ij, +}+F_{ij,-} {\bm \tau}_i \cdot {\bm \tau}_j \ ,
\end{equation}
where the $F_{ij,\pm}$ are related to the physical amplitudes
for $p$$p$ and $p$$n$ scattering (see below).  The invariant energy
$\sqrt{s_{iA}}$ is determined as follows~\cite{Schiavilla05a}.  Nucleon
$A$ denotes the knocked-out nucleon with momentum ${\bf p}_A$=${\bf p}$
and energy $E_A$=$E$ (${\bf p}$ and $E$ are the momentum and energy of
the outgoing proton in the lab frame), while nucleons $1, \dots, A-1$,
making up the bound cluster ($d$ or $t$), have momenta ${\bf p}_1, \dots,
{\bf p}_{A-1}$, with ${\bf p}_1+\dots+{\bf p}_{A-1}$=${\bf p}_f$ (${\bf p}_f$
is the momentum of the recoiling cluster in the lab frame).  The invariant
energy $\sqrt{s_{iA}}$, $i$=$1,\dots, A-1$, is obtained from
\begin{eqnarray}
s_{iA} &=& (E_i + E_A)^2-({\bf p}_i +{\bf p}_A)^2  \nonumber \\
      &\simeq & 2\, m^2+2\, E \, \sqrt{ {\bf p}_f^2/(A-1)^2+ m^2 }-
2\, {\bf p} \cdot {\bf p}_f/(A-1) \ ,
\label{eq:kin}
\end{eqnarray}
where in the second line the nucleons $1,\dots, A-1$
in the recoiling cluster are assumed to share its momentum equally,
${\bf p}_i \simeq {\bf p}_f/(A-1)$.  The momenta of nucleon $A$
and nucleon $i$, $i$=$1, \dots, A-1$, after rescattering are
${\bf p}-{\bf k}$ and ${\bf p}_f/(A-1)+{\bf k}$.  The $A-2$
spectator nucleons ($j \ne i$) have each momentum ${\bf p}_f/(A-1)$.
The pair $iA$ \lq\lq rescattering frame\rq\rq we refer to in the
following is defined as that in which nucleon $A$ and nucleon $i$
have initial momenta ${\bf p}$ and ${\bf p}_f/(A-1)$ and final
momenta ${\bf p}-{\bf k}$ and ${\bf p}_f/(A-1)+{\bf k}$, respectively.

We adopt the notation of Ref.~\cite{Schiavilla05a} and parameterize
the $N$$N$ scattering amplitude in the c.m.~frame as
\begin{equation}
(2 {\rm i}\, \overline{p})^{-1}\,
\overline{F}^{\, NN}_{ij}(\,\overline{{\bf k}},s)
=\sum_{m=1}^5 \overline{F}^{\, NN}_m(\overline{{\bf k}}^{\, 2}\!,s)
\overline{O}^{\, m}_{ij} \ ,
\label{eq:fnn}
\end{equation}
where $\overline{\bf p}$ and $\overline{\bf p}^{\, \prime}$
denote the initial and final nucleon momenta, respectively,
the $\overline{F}^{\, NN}_m$'s are functions of the invariant
energy $\sqrt{s}$ and momentum transfer $\overline{\bf k}^{\, 2}$
(with $\overline{\bf k}=\overline{\bf p}-\overline{\bf p}^{\, \prime}$),
and the five operators $\overline{O}^{\, m}_{ij}$, including central,
single and double spin-flip terms, are those listed in Eq.~(3.11) of
Ref.~\cite{Schiavilla05a}.  The overline is to indicate that the
quantities above are in the c.m.~frame.

In Ref.~\cite{Schiavilla05a} we used for the functions
$\overline{F}^{\, NN}_m$ the Gaussian parameterizations obtained
by Wallace in 1981~\cite{Wallace81}.  In the present work, instead,
we derive them from the SAID analysis~\cite{Arndt00,Arndt07,SAIDdata} of $N$$N$ elastic
scattering data from threshold up to lab kinetic energies of 3 GeV ($pp$) and 1.3 GeV ($pn$).
In Appendix~\ref{app:NNSAIDtoWallace} we discuss how the Wallace form of the amplitudes
is obtained from the SAID helicity amplitudes.

Once the amplitude in Eq.~(\ref{eq:fnn}) has been determined
in the c.m.~frame, it is necessary to boost it to the rescattering
frame.  This is carried out with the procedure described in
Refs.~\cite{Schiavilla05a,McNeil83}, which consists of two steps.
First, we introduce an invariant representation of the amplitude,
\begin{equation}
{\cal F}^{NN}_{ij}=\sum_{m=1}^5 {\cal F}_m^{NN}(s,t) \Lambda_{ij}^m \ , \label{eq:fermi_invariants}
\end{equation}
where the five operators $\Lambda_{ij}^{m=1,\dots,5}$ are $1\, ,
\, \gamma_i^\mu \, \gamma_{j,\mu} \, ,
\, \sigma_i^{\mu\nu}\, \sigma_{j,\mu\nu} \, ,
\, \gamma_i^5 \, \gamma_j^5 \, ,
\, \gamma^5_i\, \gamma_i^\mu\, \gamma^5_j\, \gamma_{j,\mu}  \ ,$
and determine the invariant functions ${\cal F}_m^{NN}(s,t)$
from the $\overline{F}^{\, NN}_m$'s in the c.m.~frame as in
Ref.~\cite{Schiavilla05a}---however, the momentum transfer
dependence of the matrix $\overline{M}_{mn}(\overline{p},
\overline{\bf k}^{\, 2})$ in Eq.~(3.16) of Ref.~\cite{Schiavilla05a},
which was neglected in that work, is now fully retained. 

Next, the scattering amplitude in the rescattering frame is
obtained from 
\begin{equation}
\chi_{\sigma_i^\prime}^\dagger \chi_{\sigma_j^\prime}^\dagger
\left[ (2 {\rm i}\, p)^{-1}\, F^{NN}_{ij}({\bf k},s)\right]
\chi_{\sigma_i} \chi_{\sigma_j}=
\overline{u}_{\sigma_i^\prime}({\bf p}-{\bf k})
\overline{u}_{\sigma_j^\prime}({\bf p}_f/(A-1)+{\bf k}) {\cal F}^{NN}_{ij}
u_{\sigma_i}({\bf p}) u_{\sigma_j}({\bf p}_f/(A-1)) \ ,
\label{eq:fres}
\end{equation}
where the $u_\sigma$ are (positive-energy) Dirac spinors with
$\overline{u}_\sigma \equiv u_\sigma^\dagger \gamma^0$, and
$\chi_\sigma$ are two-component Pauli spinors.  In practice, the
dependence upon ${\bf p}_f/(A-1)$ in the
spinors of particle $j$ is neglected (in this limit,
the rescattering and lab frames for the interacting $N$$N$ pair
coincide).  This is justified as long as $p_f/(A-1)$ is not
too large relative to $p$, the momentum of the fast ejected
proton, a condition satisfied at low missing momenta $p_f$
in the experiments of Refs.~\cite{Rvachev05,Reitz04,Strauch03}.
The resulting $F^{NN}_{ij}({\bf k},s)$ has central, single and
double spin-flip terms, and is given explicitly in Appendix~\ref{app:flab}.

Finally, carrying out the (two-dimensional) Fourier transform
in Eq.~(\ref{eq:prof}) leads to the profile operator
\begin{eqnarray}
\Gamma_{ij}({\bf b};s)&=&\Gamma^{(1)}_{ij}(b;s)
+\Gamma^{(2)}_{ij}(b;s)\, {\bm \sigma}_i \cdot {\bm \sigma}_j
+\left[ \Gamma^{(3)}_{ij}(b;s)\,{\bm \sigma}_i
+\Gamma^{(4)}_{ij}(b;s)\,{\bm \sigma}_j \right]
\cdot {\bf b}\times \hat{\bf p} \nonumber \\
&+& \Gamma^{(5)}_{ij}(b;s)\, {\bm \sigma}_i \cdot {\bf b} \,\, 
{\bm \sigma}_j \cdot {\bf b}
+ \Gamma^{(6)}_{ij}(b;s)\, {\bm \sigma}_i \cdot \hat{\bf p} \,\, 
                     {\bm \sigma}_j \cdot \hat{\bf p}
+ \Gamma^{(7)}_{ij}(b;s)\, {\bm \sigma}_i \cdot {\bf b} \,\, 
                     {\bm \sigma}_j \cdot \hat{\bf p} \nonumber \\
&+& \Gamma^{(8)}_{ij}(b;s)\, {\bm \sigma}_i \cdot \hat{\bf p} \,\,
                       {\bm \sigma}_j \cdot {\bf b} \ ,
\label{eq:gop}
\end{eqnarray} 
where the isospin-dependent operators $\Gamma^{(m)}_{ij}$, $m=1,\dots,8$,
are given by
\begin{equation}
\Gamma^{(m)}_{ij}(b;s)=\Gamma^{(m)}_+(b;s)
+\Gamma^{(m)}_-(b;s)\, {\bm \tau}_i \cdot {\bm \tau}_j \ .
\end{equation}
The profile functions $\Gamma^{(m)}_{\pm}$ are related to
those corresponding to $p$$p$ and $p$$n$ elastic scattering, obtained in
Appendix~\ref{app:flab}, via
\begin{equation}
\Gamma^{(m)}_{\pm}=\left(  \Gamma^{(m)}_{pp}\pm \Gamma^{(m)}_{pn} \right)\!/2\ .
\end{equation}

\subsection{Optical potential}
\label{sec:opt}

To describe FSI effects in the $^4$He($e,e^\prime p$)$^3$H
and $^4$He($\vec{e},e^\prime \vec{p}\,$)$^3$H reactions, we also use 
an optical potential~\cite{Schiavilla05b,vanOers82,Schiavilla90}.
In this case, the $p\, ^3$H wave function reads 
\begin{equation}
\psi^{(-)}_{{\bf k}\sigma;\sigma_3}(p+^3\!{\rm H};{\rm OPT})=
\frac{ {\rm e}^{ {\rm i}({\bf p}+{\bf p}_3)\cdot {\bf R}_{1\dots4}}}
{\sqrt{4}} \sum_{\cal P} \epsilon_{\cal P}
\Big[ \eta_{{\bf k}\sigma}^{(-)}(i;p) \phi_{\sigma_3}(jkl;^3\!{\rm H})
      +\eta_{{\bf k}\sigma}^{(-)}(i;n) \phi_{\sigma_3}(jkl;^3\!{\rm He}) \Big] \ ,
\end{equation}
where $\sigma$ and $\sigma_3$ are the spectator nucleon and bound
cluster spin projections, ${\bf k}$ and ${\bf p}+{\bf p}_3$ are their
relative and total momenta, respectively.

The spectator wave functions $\eta(i;p/n)$ are obtained from the
linear combinations $[\eta(i;T=1)$$+/-$$\eta(i;T=0) ]/2$, where
$T$=0,1 denotes the total isospin of the 1+3 clusters.  The
latter are taken to be the scattering solutions of a Schr\"odinger
equation containing a complex, energy-dependent optical potential
of the form
\begin{equation}
v^{\rm opt}_T(T_{\rm rel})= [ v^c(r;T_{\rm rel})+(4T-3)v^{c\tau}(r;T_{\rm rel})]
                           +[ v^b(r;T_{\rm rel})+(4T-3)v^{b\tau}(r;T_{\rm rel})]\,
{\bf l}\cdot {\bf s} \ ,
\end{equation}
where $T_{\rm rel}$ is the relative energy between clusters $i$ and
$j$$k$$l$, and ${\bf l}$ and ${\bf s}$ are the orbital and spin angular
momenta of nucleon $i$, respectively.  The imaginary part of $v^{\rm opt}_T$
accounts for the loss of flux in the $p\, ^3$H and $n\, ^3$He states due
to their coupling to the $dd$, three- and four-body breakup channels of
$^4$He.  Note that the $n$+$^3$He component in the scattering wave function
$\psi^{(-)}(p+^3\!{\rm H})$ vanishes unless the isospin-dependent
(charge-exchange) terms in $v^{\rm opt}$ are included.  In the results
presented in Sec.~\ref{sec:res}, all partial waves are retained in the
expansion of $\eta(i;T)$, with full account of interaction effects in those
with relative orbital angular momentum $l \leq 17$.  It has been explicitly
verified that the numerical importance of FSI in higher partial waves is
negligible.

The central $v^c$ and $v^{c\tau}$, and spin-orbit $v^b$ and $v^{b\tau}$
terms have standard Woods-Saxon and Thomas functional forms.  The
parameters of $v^c$ and $v^b$ were determined by fitting $p+^3{\rm H}$
elastic cross section data in the lab energy range $T_{\rm lab}$=(160--600) MeV,
see Ref.~\cite{vanOers82} for a listing of their values.  The parameters of the
$v^{c\tau}$ and $v^{b\tau}$ terms have been constrained by fitting
$p+^3{\rm H} \rightarrow n+^3{\rm He}$ charge-exchange cross section data
at $T_{\rm lab}$=57 MeV and 156 MeV~\cite{Schiavilla90} and the induced polarization
$P_y$ measured in the $^4$He($e,e^\prime\vec{p}\, )^3$H reaction~\cite{Malace:2011}. 
The charge-exchange central term has a real part
given by $[7.60-0.033\, T_{\rm lab}({\rm MeV})]$ MeV with radius and
diffuseness of 1.2 fm and 0.15 fm and an imaginary part given by $[0.893-0.0025\, T_{\rm lab}({\rm MeV})]$
MeV with radius and diffuseness of 1.8 fm and 0.2 fm,
while the charge-exchange spin-orbit term is taken to be purely real, with
a depth parameter depending logarithmically on $T_{\rm lab}$,
$[-15.0+1.5\, {\rm log}\, T_{\rm lab}({\rm MeV})]$ in MeV, and with radius and
diffuseness having the values 1.2 fm and 0.15 fm, respectively
(note that in Ref.~\cite{Schiavilla05b} the sign of the depth
parameter of this term had been reported erroneously with the
opposite sign).
\section{Calculation}
\label{sec:calc}

In this section we give, for completeness, a brief summary of those aspects
of the calculations, relating to the bound cluster wave functions, nuclear
electromagnetic current, and Monte Carlo methods used in evaluating the
matrix elements, which have already been reviewed in considerable detail
in Refs.~\cite{Schiavilla05a,Schiavilla05b} and references therein.

The bound states of the three- and four-nucleon systems are represented
by variational wave functions, obtained with the hyperspherical-harmonics
(HH) technique~\cite{Viviani05a} from a realistic Hamiltonian consisting of
the Argonne $v_{18}$~\cite{Wiringa95} and Urbana-IX~\cite{Pudliner95}
(AV18/UIX) potentials.  These potentials and the resulting wave functions
have been shown to account successfully at a quantitative level for a wide
variety of three- and four-nucleon properties, such as binding energies
and charge radii~\cite{Viviani05a}.

The nuclear electromagnetic current includes one- and two-body components.
The one-body operators, listed in Ref.~\cite{Schiavilla05a}, are derived from an
expansion of the covariant single-nucleon current~\cite{Jeschonnek98}.
The two-body operators used in the present work are discussed in the review
paper~\cite{Carlson98} (and references therein).  The leading terms are
derived from the static part of the AV18 potential, which is assumed to
be due to exchanges of effective pseudo-scalar ($\pi$-like) and vector
($\rho$-like) mesons.  The corresponding charge and current operators
are constructed from non-relativistic reductions of Feynman amplitudes with the
$\pi$-like and $\rho$-like effective propagators projected out of the central, spin-spin
and tensor components of the AV18.  Additional (short-range) currents
result from minimal substitution in its momentum-dependent components.
These charge and current operators contain no free parameters, and
their short-range behavior is consistent with that of the AV18.  The (purely
transverse) two-body currents associated with $M1$-excitation of $\Delta$
resonances in the intermediate state, and from $\rho\pi\gamma$ and
$\omega\pi\gamma$ transition mechanisms are
also included.  As documented in Refs.~\cite{Carlson98,Carlson02,Marcucci05},
these charge and current operators reproduce quite well a variety of few-nucleon
electromagnetic observables, ranging from elastic form factors to low-energy
radiative capture cross sections to the quasi-elastic response in inclusive
$(e, e^\prime)$ scattering at intermediate energies.

The H\"ohler parameterization~\cite{Hohler76} is used for the
electromagnetic form factors of the nucleon.  In the analysis
of the $^4$He($\vec{e},e^\prime \vec{p}\, $)$^3$H experiment, however,
at the highest $Q^2$ values of 1.6 (GeV/c)$^2$ and 2.6 (GeV/c)$^2$
the proton electric and magnetic form factors are taken from the
parameterization obtained in Ref.~\cite{Brash02} by fitting $G_{Mp}$
data and the ratio $G_{Ep}/G_{Mp}$ recently measured at JLab~\cite{Jones00}.

Finally, the numerical evaluation of the relevant matrix elements is carried
out by a combination of Monte Carlo methods and standard quadrature
techniques, described for the case of $A$=3 in Ref.~\cite{Schiavilla05a}.
This hybrid approach is easily generalized to the $A$=4 case: indeed, it
was already used in the calculations reported in Ref.~\cite{Schiavilla05b}.
The resulting predictions are numerically \lq\lq exact\rq\rq, apart
from small statistical errors due to the Monte Carlo integration, and
therefore suffer from no further approximations beyond those inherent
to the treatment of FSI and nuclear electromagnetic currents.
\section{Observables}\label{sec:observables}
For clarity we briefly recap the observables of interest for this calculation. 
More details can be found in Refs.~\cite{Schiavilla05a,Schiavilla05b} for observables
relevant to the \heedp~and \heedpfour~reactions, respectively.

The five-fold differential cross section for the $^Ai(e,e'p)^{(A-1)}f$ process is given as
\begin{equation}\label{eq:DSG_He3}
  \frac{d^5\sigma}{dE_{e}^\prime d\Omega_{e}^\prime d\Omega} = p\, E\, \sigma_{\rm Mott}\, 
  f_{\rm rec}\frac{m}{E}\frac{m_f}{E_f}
  \left[ v_{L}R_L + v_{T} R_{T} + v_{LT}\, R_{LT}\cos(\phi) + v_{TT}R_{TT}\cos(2\phi) \right],
\end{equation}
where $E_{e}^\prime$ is the energy of the final electron, 
$\Omega_{e}^\prime$ and $\Omega$ are, respectively, the solid angles of the final electron
and ejected proton, $m_f$ is the rest mass of the ($A$--1)-cluster, 
${\bf p}$ and $E$ (${\bf p}_f$ and $E_f$) are the momentum and energy of the proton (($A$--1)-cluster),
$\phi$ is the angle between the electron scattering plane and the plane defined by ${\bf q}$ and ${\bf p}$,
and the recoil factor is defined by its inverse
\begin{equation}
  f_{\rm rec}^{-1} = \left| 1 - \frac{p_f\, E}{p\, E_f} {\hat{\bf p}} \cdot {\hat{\bf p}}_f \right|.
\end{equation}
For a derivation of Eq.~(\ref{eq:DSG_He3}), the definition of $\sigma_{\rm Mott}$ and of the
(standard) electron kinematic factors, 
$v_{\alpha}$, where $\alpha = L, T, LT, TT$, see Ref.~\cite{Raskin89}. 
The nuclear response functions are given in Ref.~\cite{Schiavilla05a}.

The longitudinal-transverse asymmetry $A_{LT}$ is obtained from the differential cross sections
\begin{align}\label{eq:A_LT}
  A_{LT} &= \frac{\sigma(\phi = 0^{\circ}) - \sigma(\phi = 180^{\circ})}
  {\sigma(\phi = 0^{\circ}) + \sigma(\phi = 180^{\circ})} \nonumber \\
  &= \frac{v_{LT}R_{LT}}{v_{L}R_{L} + v_{T}R_{T} + v_{TT}R_{TT}},
\end{align}
where $\sigma(\phi)$ represents the differential cross section in Eq.~(\ref{eq:DSG_He3}).

In parallel kinematics, where the electron three-momentum transfer ${\bf q}$
and the missing momentum ${\bf p}_m$ (defined as ${\bf p}_m=-{\bf p}_f={\bf p}-{\bf q}$)
are parallel, the polarization
transfers $P_x^\prime$ and $P_z^\prime$ are given by
\begin{equation}
P_x^\prime= \frac{v_{LT^\prime}R^t_{LT^\prime}}{v_{L}R_{L} + v_{T}R_{T}} \ , \qquad
P_z^\prime=\frac{v_{TT^\prime}R^l_{TT^\prime}}{v_{L}R_{L} + v_{T}R_{T}} \ ,
\end{equation}
where the response functions $R^t_{LT^\prime}$ and $R^l_{TT^\prime}$ and electron
kinematical factors $v_{LT^\prime}$ and $v_{TT^\prime}$ read~\cite{Picklesimer89}
\begin{eqnarray}
R^t_{LT^\prime}&=&2\,\sqrt{2}\sum_{m_3}  {\rm Im} \Big[
\langle p+^3\!{\rm H}; +\hat{\bf x},m_3\!\mid \rho (q\hat{\bf z}) \mid ^4\!{\rm He}\rangle\, 
\langle p+^3\!{\rm H}; +\hat{\bf x},m_3\!\mid  j_y (q\hat{\bf z}) \mid ^4\!{\rm He}\rangle^* \Big] \ , \\
R^l_{TT^\prime}&=&2 \sum_{m_3}  {\rm Im} \Big[
\langle p+^3\!{\rm H}; +\hat{\bf z},m_3 \!\mid j_x (q\hat{\bf z}) \mid ^4\!{\rm He}\rangle\, 
\langle p+^3\!{\rm H}; +\hat{\bf z},m_3\!\mid  j_y (q\hat{\bf z}) \mid ^4\!{\rm He}\rangle^* \Big] \ ,\\
&&v_{LT^\prime}=\frac{1}{\sqrt{2}} \,  \frac{Q^2}{q^2} {\rm tan}( \theta_e/2)
\ ,\qquad v_{TT^\prime}= {\rm tan}( \theta_e/2) \sqrt{ \frac{Q^2}{q^2}+{\rm tan}^2( \theta_e/2)} \ .
\end{eqnarray}
and $\theta_e$ and $Q^2=q^2-\omega^2$ are, respectively,
the electron scattering angle and four-momentum transfer.
In the above equations, \hbox{$\mid\! ^4{\rm He}\rangle$} represents the $^4$He ground state, while
\hbox{$\mid\! p+^3\!{\rm H}; +\hat{\bf x},m_3 \rangle$} and
\hbox{$\mid\! p+^3\!{\rm H}; +\hat{\bf z},m_3 \rangle$}
represent the $p+^3\!{\rm H}$ final scattering states with the proton spin projection along
either the $\hat{\bf x}$ or the $\hat{\bf z}$ directions, respectively, and with the $^3$H in spin
projection $m_3$.  The momentum transfer ${\bf q}$ has been taken along the $\hat{\bf z}$
direction, which also defines the quantization axis of the proton and $^3$H spins.
Then, the $\mid\!\! p+^3\!{\rm H}; +\hat{\bf x},m_3 \rangle$ state, having the
proton polarized in the $\hat{\bf x}$ direction, is written as
\begin{equation}
\mid\!\! p+^3\!{\rm H}; +\hat{\bf x},m_3 \rangle = 
\frac{1}{\sqrt{2}} \mid\!\! p+^3\!{\rm H}; +\hat{\bf z},m_3 \rangle 
+ \frac{1}{\sqrt{2}} \mid\!\! p+^3\!{\rm H}; -\hat{\bf z},m_3 \rangle \ ,
\end{equation}
and the amplitudes
$\langle p+^3\!{\rm H}; \pm\, \hat{\bf z},m_3\!\mid  O (q\hat{\bf z}) \mid ^4\!{\rm He}\rangle$
are calculated for all possible combinations of proton and $^3$H spin
projections and of transition operators $O(q\hat{\bf z})$ with the methods
discussed in the previous section.

Lastly, the induced polarization $P_y$ is defined as
\begin{equation}
P_y=\frac{v_{LT}\, \Delta R_{LT}}{v_{L}R_{L} + v_{T}R_{T}}
\end{equation}
where the $\Delta R_{LT}$ response function is defined as
\begin{eqnarray}
\Delta R_{LT}&=&2\,\sqrt{2}\sum_{m_3}  {\rm Re} \Big[
\langle p+^3\!{\rm H}; +\hat{\bf y},m_3\!\mid \rho (q\hat{\bf z}) \mid ^4\!{\rm He}\rangle\, 
\langle p+^3\!{\rm H}; +\hat{\bf y},m_3\!\mid  j_x (q\hat{\bf z}) \mid ^4\!{\rm He}\rangle^* \nonumber\\
&&-\langle p+^3\!{\rm H}; -\hat{\bf y},m_3\!\mid \rho (q\hat{\bf z}) \mid ^4\!{\rm He}\rangle\, 
\langle p+^3\!{\rm H}; -\hat{\bf y},m_3\!\mid  j_x (q\hat{\bf z}) \mid ^4\!{\rm He}\rangle^* \Big] \ ,
\end{eqnarray}
and in the states \hbox{$\mid\! p+^3\!{\rm H}; \pm\, \hat{\bf y},m_3 \rangle$}
the proton polarization is along the $\pm \, \hat{\bf y}$ direction (note that in
parallel kinematics, the proton and electron scattering planes concide, and
are taken here as the $xz$-plane).

\section{Results} \label{sec:res}
In this section we compare the results of our calculations to experimental data.
In addition we compare various model-dependent effects, and discuss
how these affect the results.
\subsection{\heedp}
\label{subsec:3he}

\begin{figure}
    \includegraphics[width=16cm]{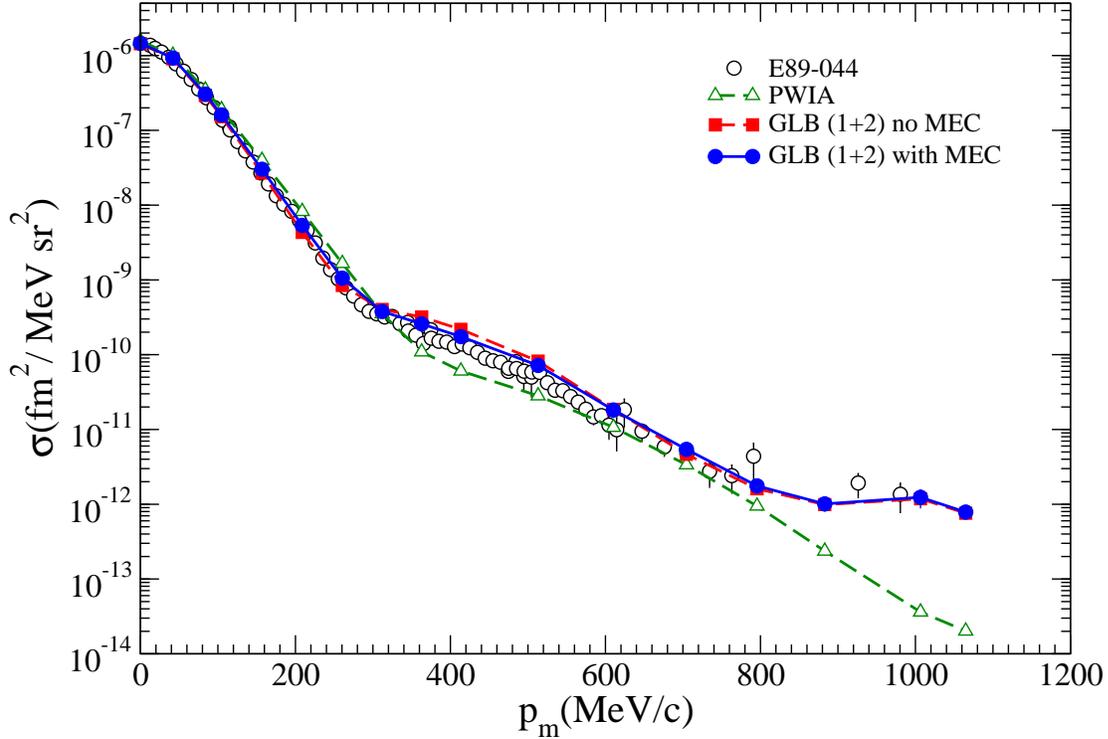}  
    \caption{(Color online) Differential cross sections for the \heedp~reaction at $\phi$=180$^{\circ}$. 
    The experimental data are compared with the plane-wave-impule-approximation (PWIA), 
    and with the full single and double rescattering 
    Glauber approximation with MEC (GLB(1+2) With MEC) and without MEC (GLB(1+2) No MEC).
    The profile operator in the Glauber approximation is derived from $NN$ scattering amplitudes
    (including central, single-spin flip and double-spin flip terms), boosted from the c.m.~frame
    to the rescattering (lab) frame.  Statistical Monte Carlo errors are smaller than the symbols, and
    lines drawn to guide the eye.}
    \label{fig:avg_DSG_kin1}
\end{figure}

\begin{figure}
    \includegraphics[width=16cm]{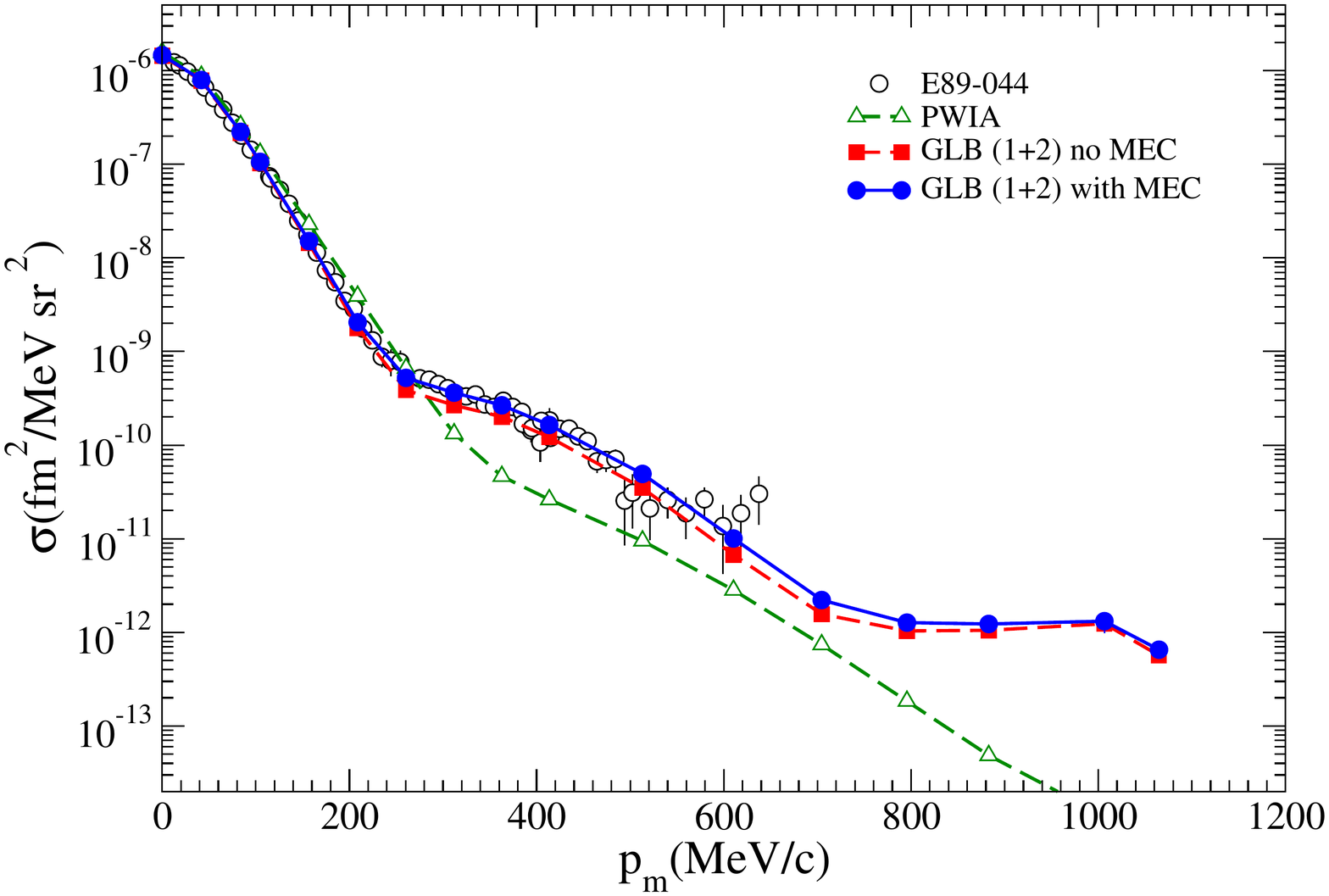}  
    \caption{(Color online) Same as Fig. \ref{fig:avg_DSG_kin1}, but at $\phi$=0$^{\circ}$.}
    \label{fig:avg_DSG_kin2}
\end{figure}

\begin{figure}
    \includegraphics[width=16cm]{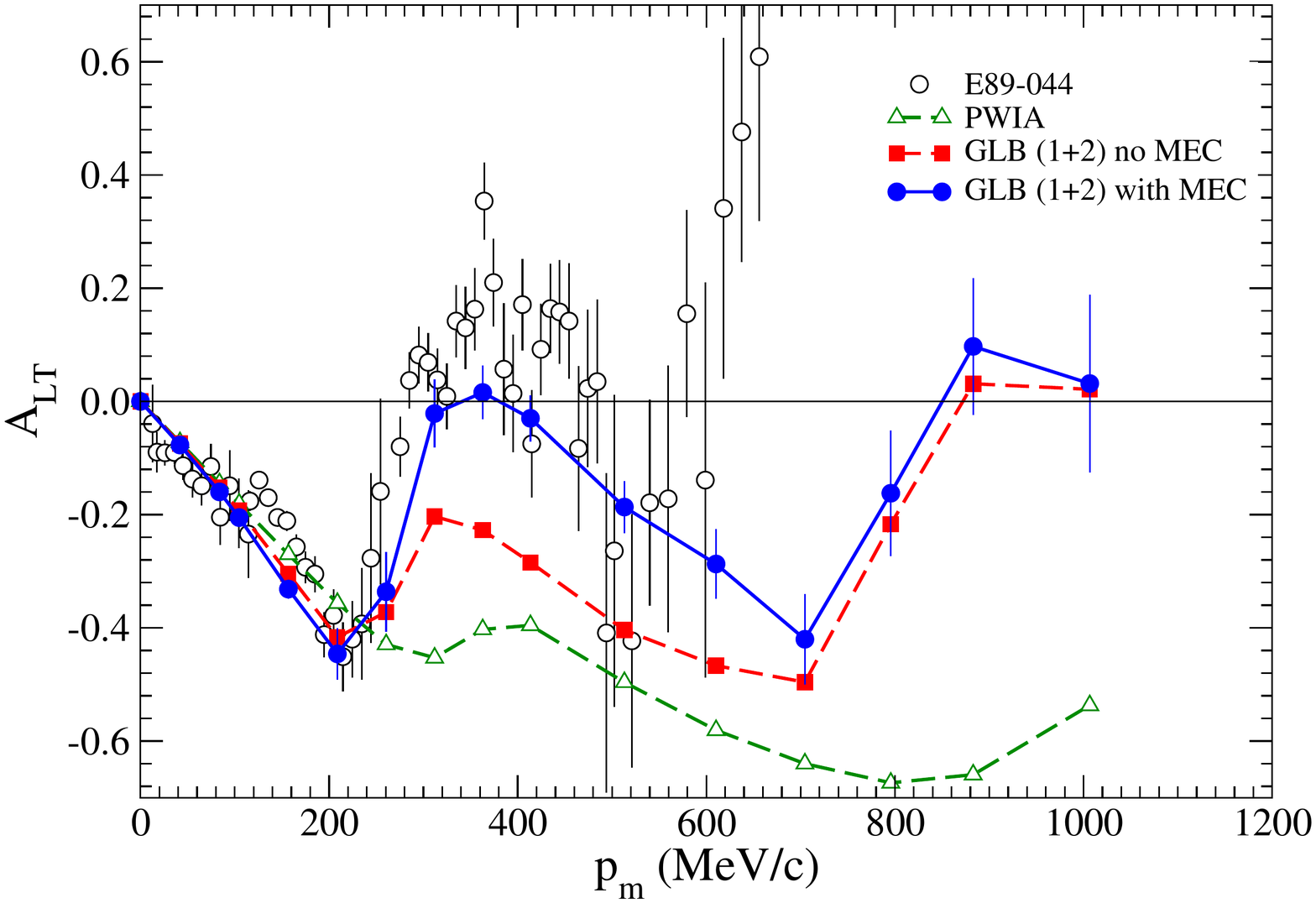}  
    \caption{(Color online) Same as Fig.~\ref{fig:avg_DSG_kin1}, but for the longitudinal-transverse asymmetry.}
    \label{fig:avg_ASY}
\end{figure}

As in Ref.~\cite{Schiavilla05a} the predicted cross section and asymmetry 
are compared with experimental data taken at JLab (E89-044)~\cite{Rvachev05}. 
For the \heedp~reaction all observables are plotted as function of the missing
momentum $p_m$.  The calculated cross sections are compared to experimental
data for $\phi$=180$^\circ$ in Fig.~\ref{fig:avg_DSG_kin1} and for $\phi$=0$^\circ$ in
Fig.~\ref{fig:avg_DSG_kin2}.  The longitudinal-transverse asymmetry is obtained
from these cross sections via Eq.~(\ref{eq:A_LT}), and its comparison to experiment
is shown in Fig.~\ref{fig:avg_ASY}. 

In Figs.~\ref{fig:avg_DSG_kin1}--\ref{fig:avg_ASY}, the curves labeled PWIA represent the
results obtained in the plane-wave impulse-approximation disregarding all FSI.  The PWIA
overpredicts the data at low $p_m$ and underpredicts them at high $p_m$.
The curves labeled ``GLB(1+2) No MEC'' represent the results obtained
in the Glauber approximation with single and double rescattering, 
but neglecting contributions from meson exchange currents (MEC).
By accounting for FSI, we note a significant improvement in describing the experimental
cross-section values. 
Inclusion of MEC contributions, curves labeled as ``GLB(1+2) With MEC'', 
further improves the comparison with the data. 
While in Figs.~\ref{fig:avg_DSG_kin1} and~\ref{fig:avg_DSG_kin2} the MEC effects appear small in comparison to the FSI, it is clear that they improve the predictions, especially
at intermediate values of missing momentum.
In Fig.~\ref{fig:avg_ASY}, where the asymmetry is shown, the effects are even more pronounced. 
We note again the inability of the PWIA to successfully account for the experimental features, 
except for very low values of missing momentum.  The structure of the data is clearly
dominated by FSI as the missing momentum is increased, and the importance of the MEC
is again notable.  Indeed at intermediate values of $p_m$ the MEC contribution is of
comparable strength as the FSI.  When calculating $A_{LT}$, we are taking a difference
between the cross sections shown in Fig.~\ref{fig:avg_DSG_kin1} and Fig.~\ref{fig:avg_DSG_kin2},
where in the former the MEC suppress the results, and in the latter the MEC enhance them.
So even though this is a small effect in the individual cross sections, it becomes
quite large when taking their difference.

\begin{figure}
    \includegraphics[width=16cm]{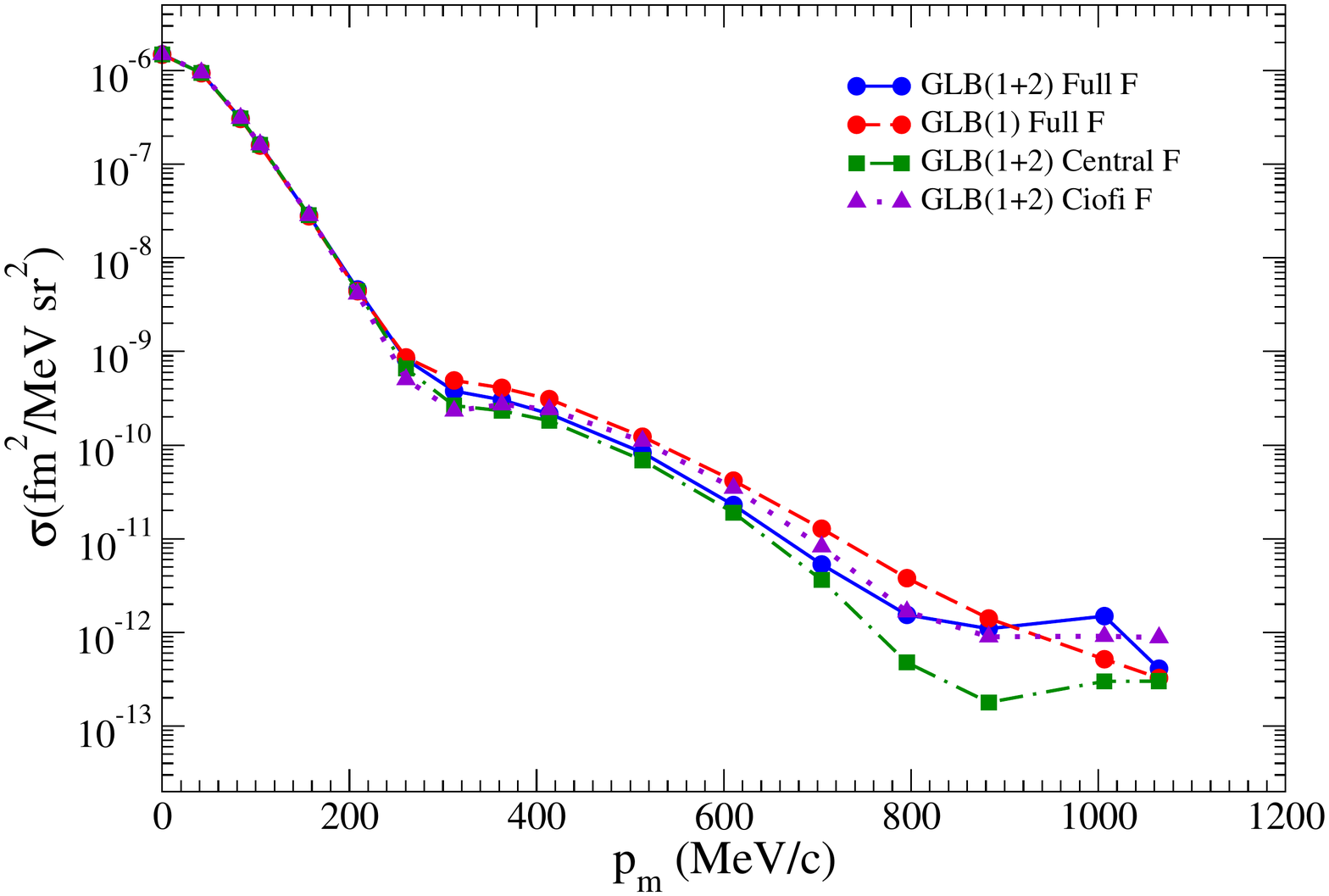}  
    \caption{(Color online) Differential cross sections for the \heedp~reaction at $\phi$=180$^{\circ}$ 
    obtained in various approximation schemes, see text for descriptions of the approximations.  Lines are
    drawn to guide the eye.}
    \label{fig:comp_DSG_kin1}
\end{figure}

\begin{figure}
    \includegraphics[width=16cm]{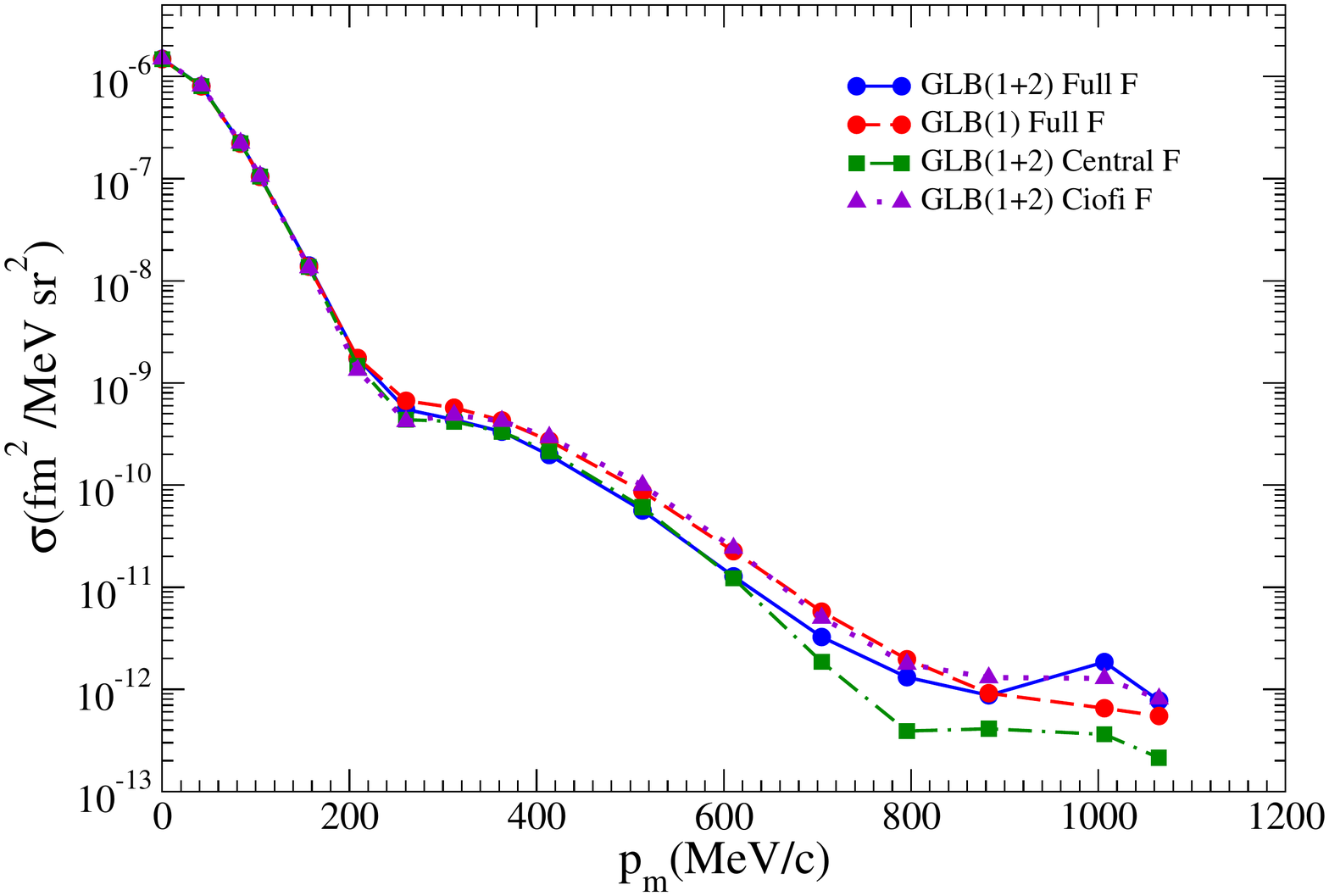}  
    \caption{(Color online) Same as Fig.~\ref{fig:comp_DSG_kin1}, but at $\phi$=0$^{\circ}$.}
    \label{fig:comp_DSG_kin2}
\end{figure}

\begin{figure}
    \includegraphics[width=16cm]{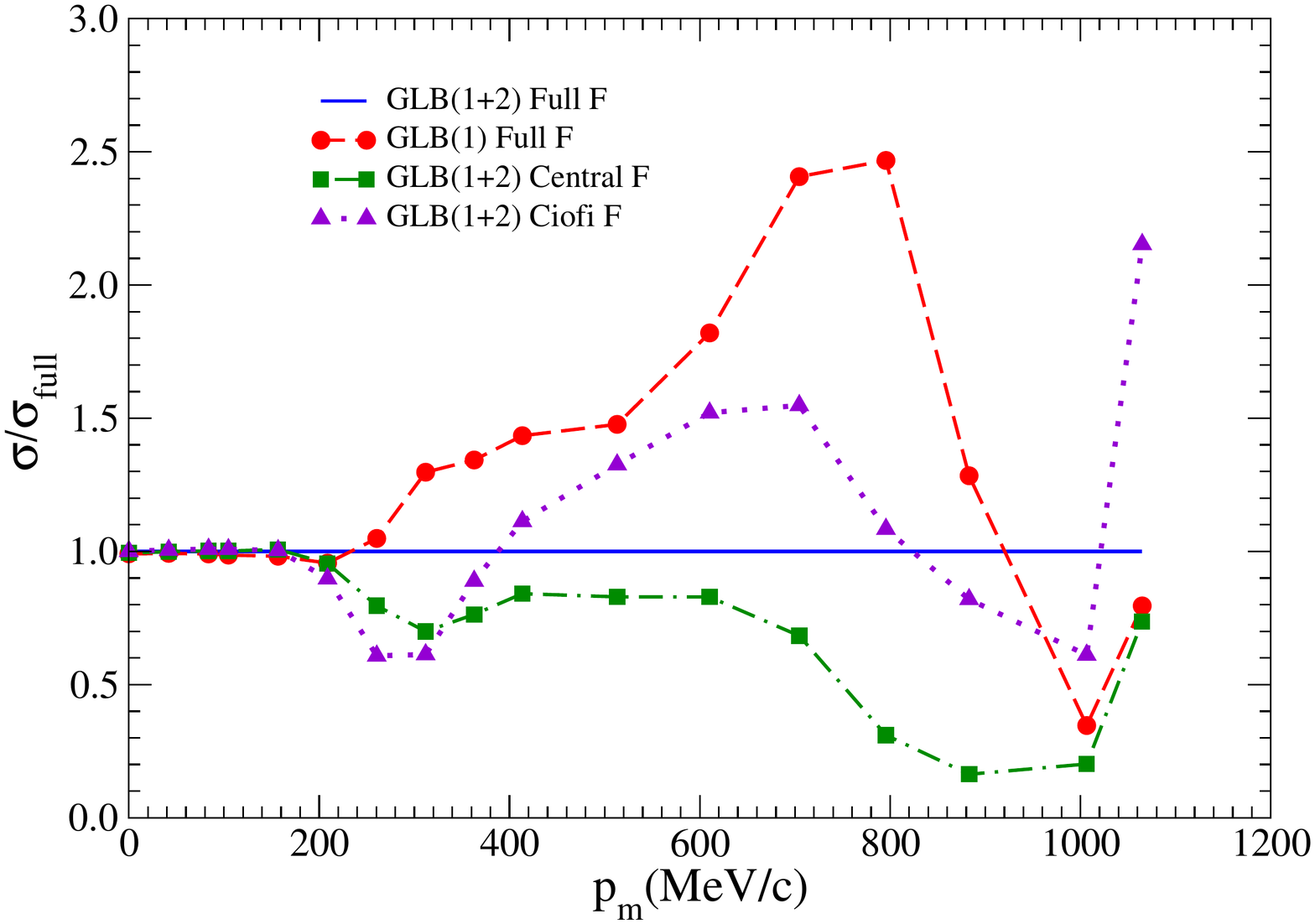}  
    \caption{(Color online) Ratios of differential cross sections
    as shown in Fig.~\ref{fig:comp_DSG_kin1}.}
    \label{fig:comp_DSGratio_kin1}
\end{figure}

\begin{figure}
    \includegraphics[width=16cm]{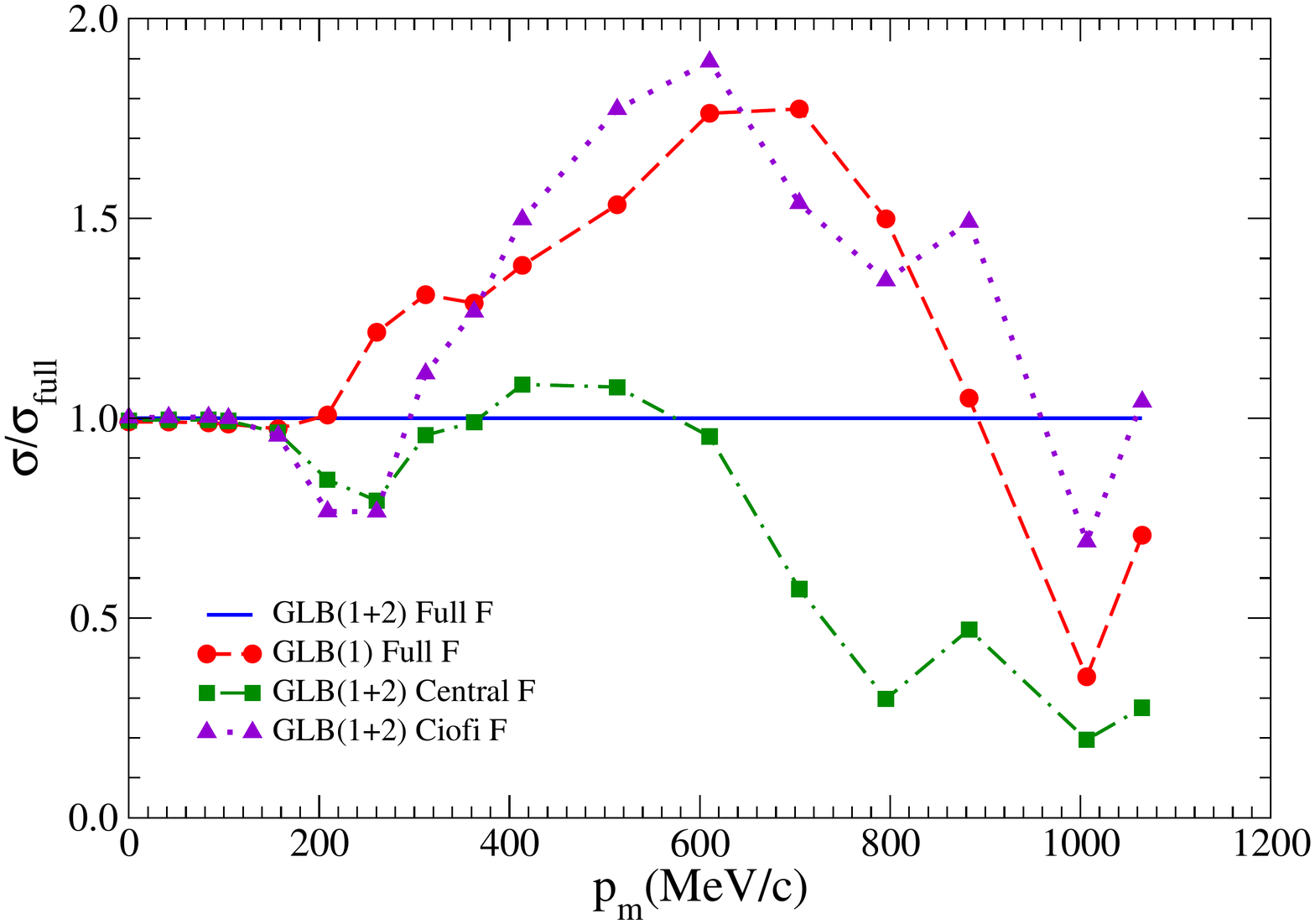}  
    \caption{(Color online) Ratios of differential cross sections as shown in Fig.~\ref{fig:comp_DSG_kin2}.}
    \label{fig:comp_DSGratio_kin2}
\end{figure}

\begin{figure}
    \includegraphics[width=16cm]{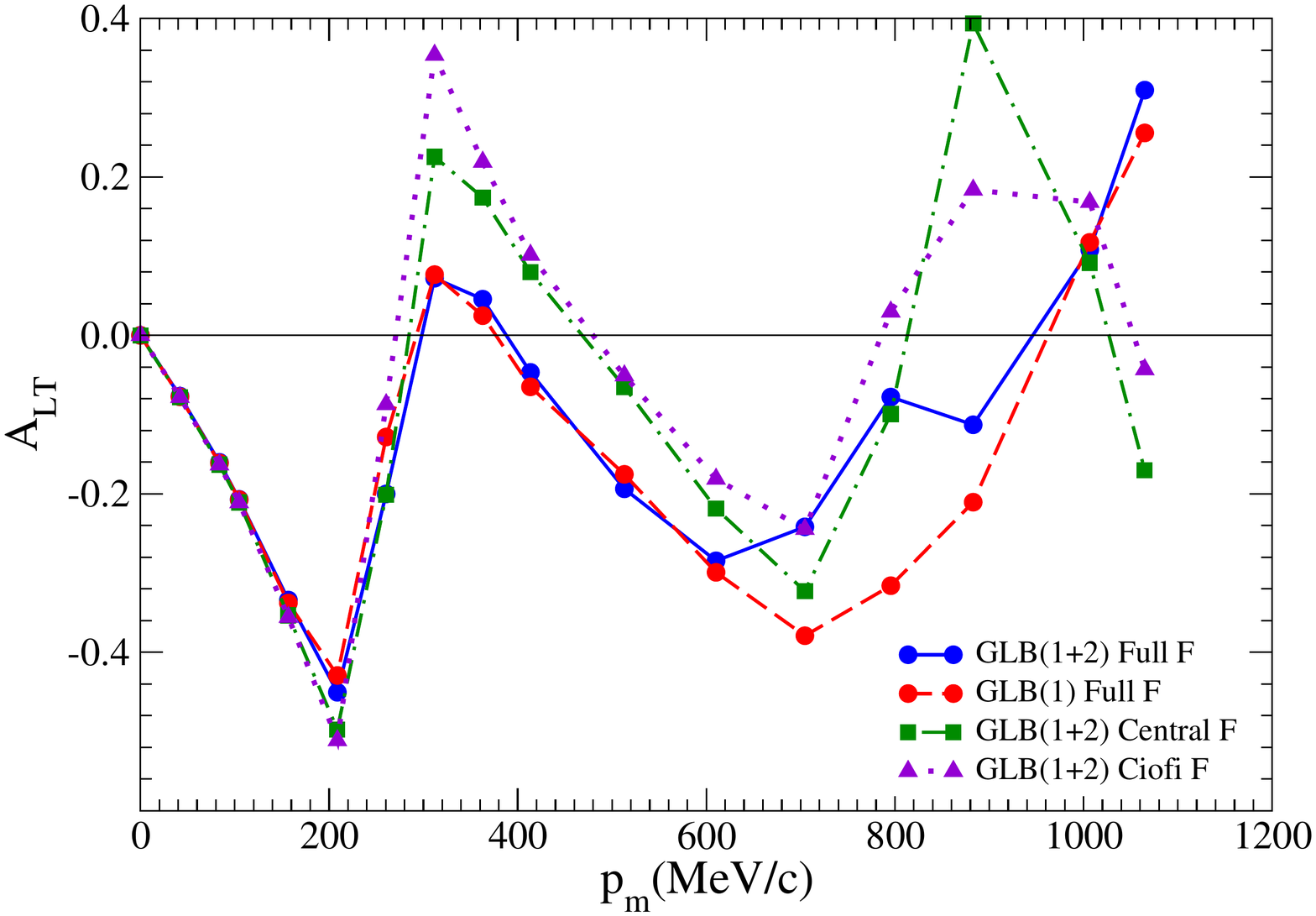}  
    \caption{(Color online) Same as Fig.~\ref{fig:comp_DSG_kin1}, but for the longitudinal-transverse asymmetry.}
    \label{fig:comp_ASY}
\end{figure}

We next want to investigate model-dependent effects due to the $NN$ scattering amplitudes. 
In order to compare the various effects, calculations were performed for a variety of cases, 
and comparisons are presented in Figs.~\ref{fig:comp_DSG_kin1}--\ref{fig:comp_ASY}.
Specifically, we are interested in quantifying the role of spin dependence in the FSI, 
which this work facilitates well, since all spin dependence is retained in the Glauber
profile operator.  These cases were all calculated for the same random walk in the Monte
Carlo integration, and are not compared to experimental data.
Since we have already investigated MEC contributions and noted their
importance in the discussion above, all results below include them.
The cases we investigate are:
\begin{enumerate}
 \item Curves labeled ``GLB(1+2) Full F'' correspond to the Glauber approximation
with single and double rescattering and include the full spin dependence in the $NN$
scattering amplitudes.
 \item Curves labeled ``GLB(1)'' include only single scattering in the Glauber approximation, but still incorporate the full spin dependence in the $NN$ scattering amplitudes.
 \item Curves labeled ``GLB(1+2) Central F'' correspond to Glauber single and double rescattering,
but with all spin dependent terms turned off in the $NN$ scattering amplitudes, that is, in
Eq.~(\ref{eq:fnn}) we set $\overline{F}^{\, NN}_m(\overline{{\bf k}}^{\, 2}\!,s)=0$ 
for $m$=2--5, so that only the central term $\overline{F}^{\, NN}_1(\overline{{\bf k}}^{\, 2}\!,s)$ contributes.
 \item Curves labeled ``GLB(1+2) Ciofi F'' correspond to using a common $NN$ parametrization given in Eq.~(\ref{eq:cioffiAmp}), which includes no explicit spin dependence.
The parameterization is described in Appendix~\ref{app:NNSAIDtoWallace}. 
It should be noted, however, that when fitting a spin independent amplitude to experimental data,
spin dependence can implicitly enter the parameterization, 
which causes some ambiguity when trying to determine its role in FSI.
\end{enumerate}

In Figs.~\ref{fig:comp_DSG_kin1} and~\ref{fig:comp_DSG_kin2} we show the
differential cross sections calculated at $\phi$= 180$^\circ$ and $\phi$=0$^{\circ}$, respectively. 
Since these are semilog plots, we also plot ratios of the various cases to the full double rescattering, 
fully spin dependent calculation, case 1. These are shown in Figs.~\ref{fig:comp_DSGratio_kin1}
and~\ref{fig:comp_DSGratio_kin2}, again for $\phi$=180$^\circ$ 
and $\phi$=0$^{\circ}$, respectively.  In Fig. \ref{fig:comp_ASY}
we plot the longitudinal-transverse asymmetry for comparison of the four cases.
In these figures we first note the necessity of including
the double rescattering in the Glauber approximation, case 2.
At all but the lowest missing momentum, where FSI are negligible, 
we see that the single scattering approximation leads to 
a significant deviation for $p_m> 200 ~\rm{MeV/c}$. 
Next we observe the effect of ``turning off'' the spin dependent contributions
the $NN$ amplitudes, case 3.  Here again we note significant deviations from
the full result.  Finally, we turn to case 4 and note similar deviations as in case 3
for $p_m \lesssim 400$ MeV/c.  However, at larger $p_m$ where
FSI effects become quite important, predictions for cases 3 and 4
differ significantly from each other---see
Figs.~\ref{fig:comp_DSGratio_kin1}--\ref{fig:comp_DSGratio_kin2}---which
can be traced back to differences
between the central amplitudes of cases 3 and 4
(see discussion in Appendix~\ref{app:NNSAIDtoWallace}).

It is interesting to point out that for the asymmetry, shown in
Fig.~\ref{fig:comp_ASY}, the effects are similar for each of the
four cases, however, we note that there is no significant deviation
for the single and double rescattering up to $p_m \approx 600~\rm{MeV/c}$.
This implies that the effects of double rescattering, so pronounced
in the differential cross section for $p_m> 200 ~\rm{MeV/c}$, cancel
when calculating the asymmetry.  This is similar to the above discussion
regarding MEC, and again is due to taking differences of cross sections,
except here the double scattering contribution increases the 
cross sections for both kinematics so when taking the difference
this increase is canceled out.
\subsection{\heedpfour}
\label{subsec:4he}
We now turn our attention to the observables calculated for the \heedpfour ~reaction. 
In this case we utilize both the Glauber description of FSI as well as an optical potential.
We begin by discussing JLab experiment E97-111, for which preliminary
data have been published in Ref.~\cite{Reitz04}---these preliminary data, which
only include statistical errors, are shown in the figures below.  The experiment
measured cross sections
for the electrodisintegration of $^4$He into $^3$H and $p$ clusters
in three different kinematic setups.  The first setup labeled CQ2, in which
the electron momentum and energy transfers
were kept fixed at $q \simeq1.43$ GeV and $\omega \simeq 0.52$ GeV,
was in quasi-perpendicular kinematics (with the missing momentum ${\bf p}_m$ close
to being perpendicular to ${\bf q}$), while the remaining two setups labeled
PY1 and PY2 were both in quasi-parallel kinematics (with
${\bf p}_m$ close to being parallel to ${\bf q}$) and both covered the
same range $ 0\lesssim p_m \lesssim 500$ MeV/c, but
the electron beam energy and scattering angle were, respectively, about 2.4 GeV and
16.9$^\circ$ in PY1 and about 3.2 GeV and 18.9$^\circ$ in PY2.

\begin{figure}
    \includegraphics[width=16cm]{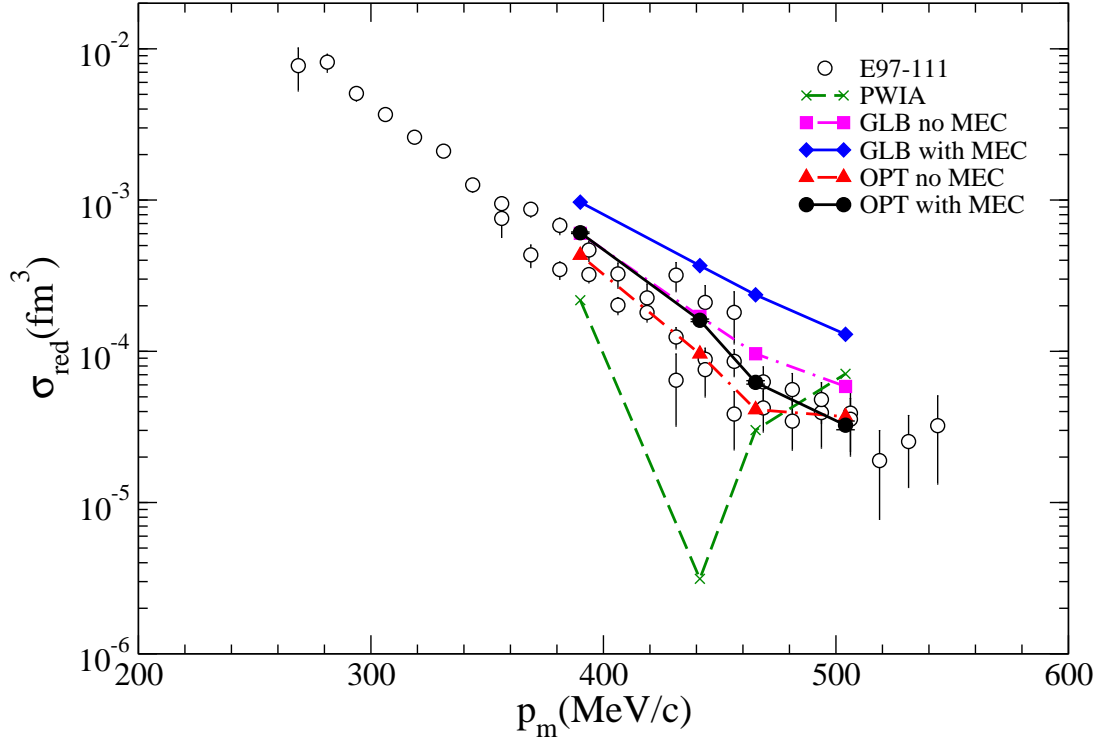}  
    \caption{(Color online) Reduced differential cross section of \heedpfour ~compared to experimental data 
    and various calculation schemes. ``CQ2'' refers to the experimental kinematics. See text for descriptions of the curves.
     Lines are drawn to guide the eye.}
    \label{fig:red_cq2}
\end{figure}

\begin{figure}
    \includegraphics[width=16cm]{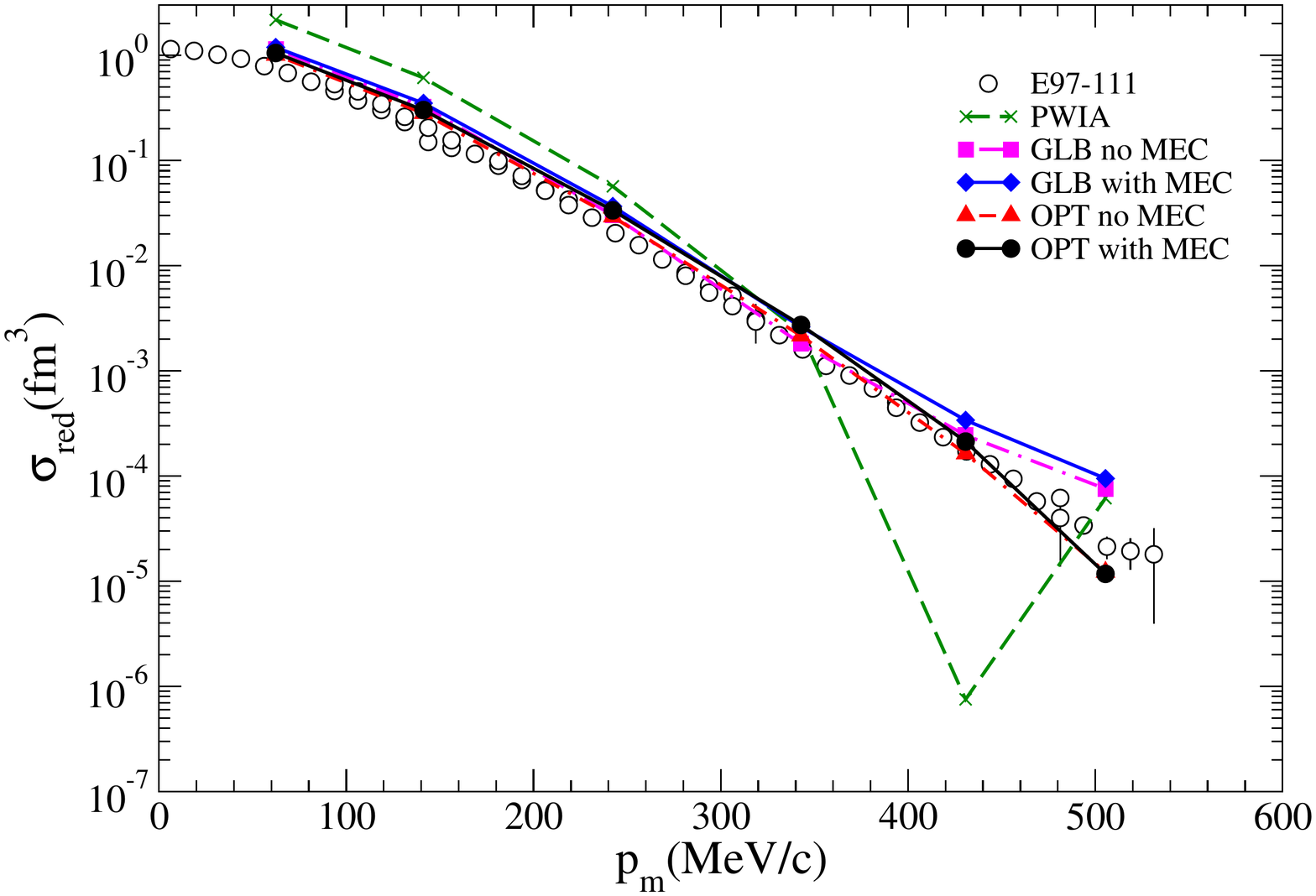}  
    \caption{(Color online) Same as Fig. \ref{fig:red_cq2} except for ``PY1'' kinematics.}
    \label{fig:red_py1}
\end{figure}

\begin{figure}
    \includegraphics[width=16cm]{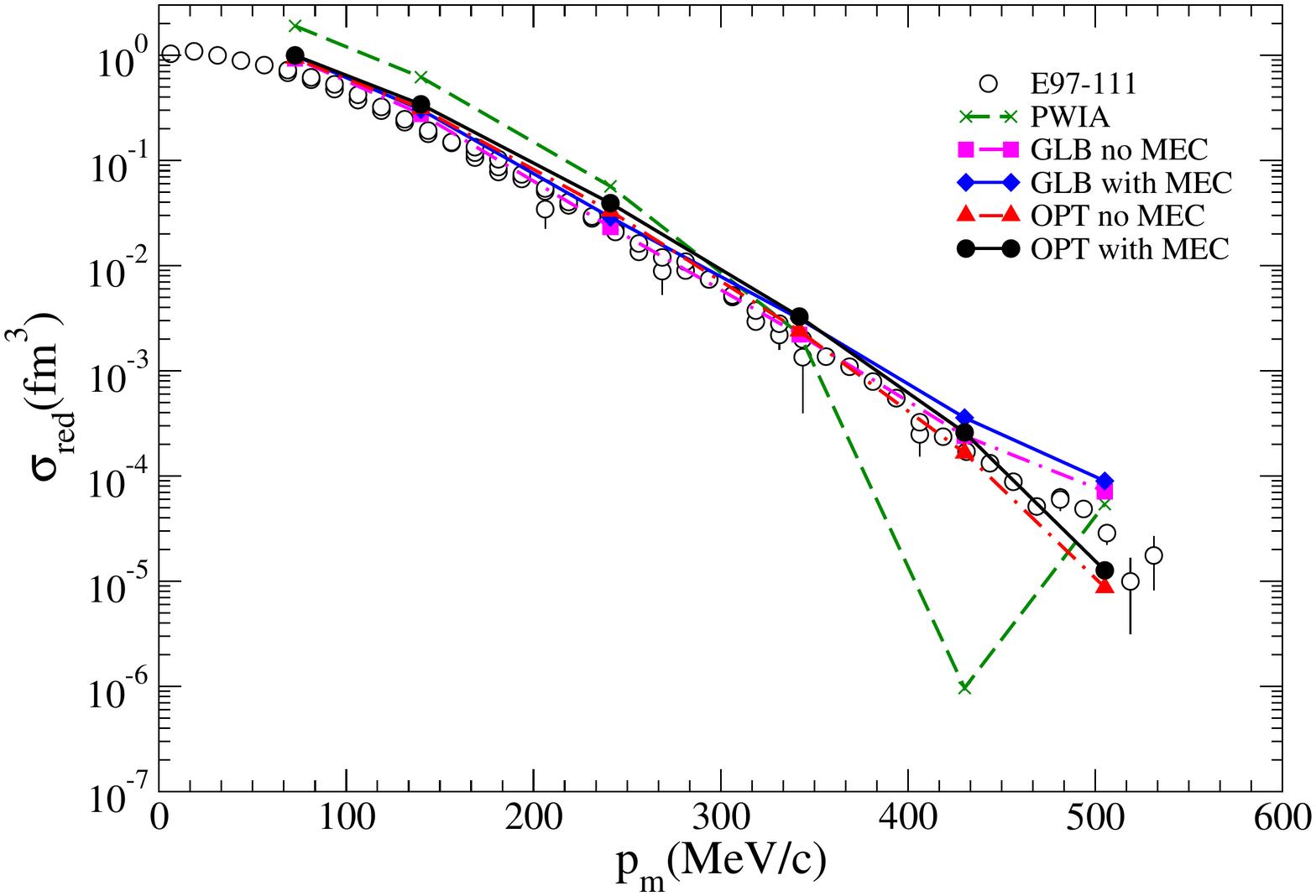}  
    \caption{(Color online) Same as Fig. \ref{fig:red_cq2} except for ``PY2'' kinematics.}
    \label{fig:red_py2}
\end{figure}

In Figs.~\ref{fig:red_cq2}--\ref{fig:red_py2} we show for both experiment and
theory the reduced cross section, defined as
\begin{equation}
\sigma^{\rm red}=\frac{1}{p\, E\, f_{\rm rec}\, \sigma_{ep}^{\rm CC1}}\,
\frac{d^5\sigma}{dE_{e}^\prime d\Omega_{e}^\prime d\Omega} \ ,
\end{equation}
where $\sigma_{ep}^{\rm CC1}$ denotes the CC1 off-shell parameterization
of the electron-proton cross section due to deForest~\cite{deForest83}.
The various curves are labeled as follows: ``PWIA'' represents the plane wave impulse
approximation, ``GLB with (no) MEC'' treats FSI in the Glauber approximation with (without) MEC,
``OPT with (no) MEC'' uses the optical potential to account for FSI with (without) MEC,
and finally the experimental data are labeled by the experiment number ``E97-111''.
We note that the calculations in the Glauber approximation
include single, double and triple rescattering (see Sec.~\ref{sec:glb}).

The three kinematic setups all cover the region of missing momentum 
close to 450 MeV/c, where the PWIA results are orders of magnitude
smaller than the data.  In PWIA the cross section is proportional
to the $p$-$^3$H cluster momentum distribution, which exhibits a node
for $p_m$ close to 450 MeV/c~\cite{Schiavilla86, *Wiringa:2013ala}.
This node is filled in by FSI contributions, which shift PWIA strength
from the low $p_m$ region to the high $p_m$ one, see
Figs.~\ref{fig:red_py1}--\ref{fig:red_py2}.  The contributions from MEC
are significant, particularly for kinematics CQ2, and increase
the cross section over the whole $p_m$ range of interest. 
The full calculations, including FSI either in the Glauber approximation
or via the optical potential and MEC contributions, are in reasonable
agreement with data for kinematics PY1 and PY2, although
they both tend to overpredict the measured cross sections at low $p_m$
(but not as severely as the PWIA calculation).  For kinematics CQ2,
the ``OPT with MEC'' calculation provides a satisfactory
description of data, while the ``GLB with MEC'' calculation leads
to cross sections which are significantly larger than the measured
values.  We note that for kinematics CQ2 the relative
kinetic energy between the proton and triton clusters is about 0.31 GeV, so
well within the range of applicability of the optical potential, which
was fitted to $p$-$^3$H scattering data up to relative kinetic
energy of 0.45 GeV (Sec.~\ref{sec:opt}).  In contrast, the proton
lab kinetic energies for this same kinematic setup are of the order of 0.46 GeV,
arguably too low for the validity of the Glauber approximation.  For the
quasi-parallel kinematics PY1 (PY2) the $p$-$^3$H relative
kinetic energies and proton lab kinetic energies are, respectively, in the
ranges 0.22--1.05 (0.44--1.55) GeV and 0.34--0.98 (0.51--1.42) GeV,
as the missing momentum increases from $\simeq 0.06$ GeV/c
to $\simeq 0.5$ GeV/c, and therefore one would expect
the treatment of FSI via the optical potential to be valid on the low side
of $p_m$ and that based on the Glauber approximation to be appropriate for
the high side of $p_m$.  In fact, the actual calculations shown in
Figs.~\ref{fig:red_py1}--\ref{fig:red_py2} indicate that the optical potential and
Glauber approximation differ significantly only beyond $p_m \gtrsim 400$ MeV/c,
with the ``OPT with MEC'' and ``GLB with MEC'' results, respectively,
underestimating and overestimating the data.
 
\begin{figure}
    \includegraphics[width=16cm]{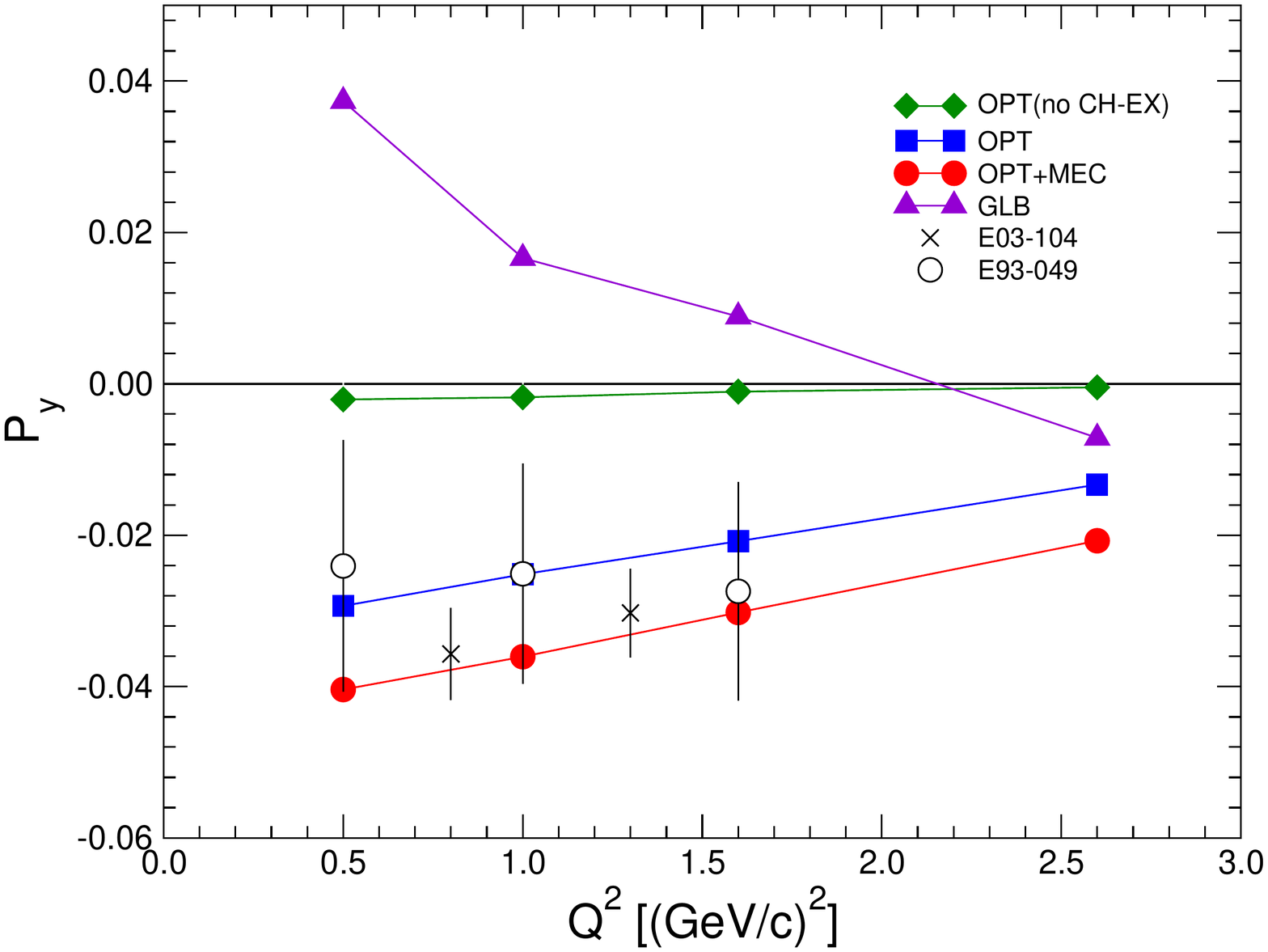}  
    \caption{(Color online) Induced polarization for $^4$He compared to experimental data.
    The optical potential is tuned to reproduce the data.
    See text for descriptions of the curves.  Lines are
    drawn to guide the eye.}
    \label{fig:py}
\end{figure}

\begin{figure}
    \includegraphics[width=16cm]{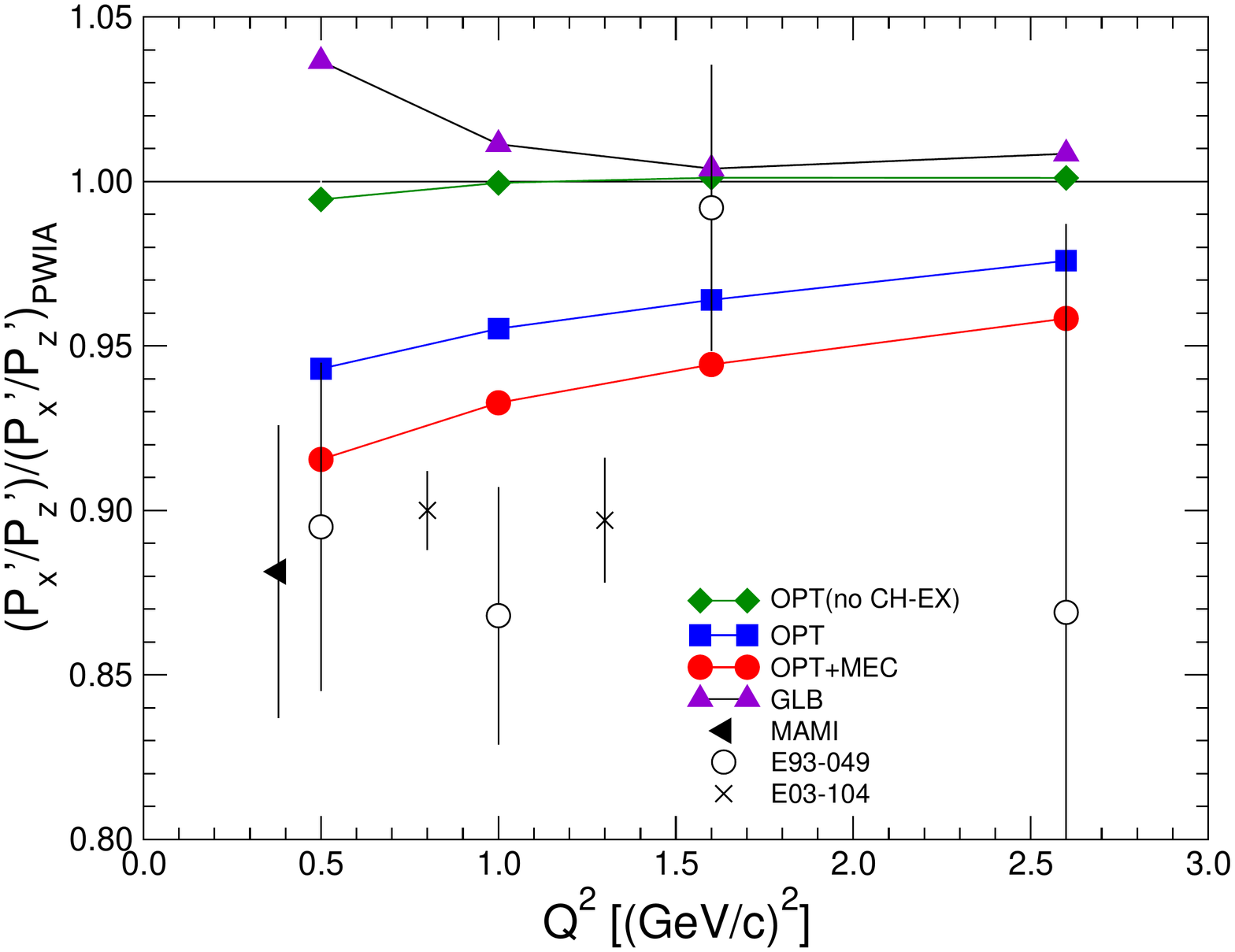}  
    \caption{(Color online) Polarization transfer for $^4$He compared to experimental data. 
    See text for descriptions of the curves.  Lines are
    drawn to guide the eye.}
    \label{fig:ratio}
\end{figure}

We now turn our attention to the polarization observables in the
$^4$He$(\vec{e},e^\prime\vec{p}\,)^3$H reaction.  We present the
induced polarization $P_{y}$ in Fig.~\ref{fig:py}, and the super-ratio
$(P^\prime_x/P^\prime_z)/(P^\prime_x/P^\prime_z)_{\rm PWIA}$ in
Fig.~\ref{fig:ratio}.

These are both plotted versus the four momentum
transfer of the virtual photon, $Q^2$.  These observables are compared
with data labeled according to the experiment.  In Fig.~\ref{fig:py}
the data labeled ``E03-104'' are from Ref.~\cite{Malace:2011}, and
``E93-049'' are from Ref.~\cite{Strauch03}.  In Fig.~\ref{fig:ratio} the
data labeled ``E03-104'' are from Ref.~\cite{Paolone10}, ``E93-049'' are
from Ref.~\cite{Strauch03}, and ``MAMI'' are from Ref.~\cite{Dieterich01}. 
When comparing to the JLab experimental data we should be mindful 
that these are averaged over the acceptance of the spectrometers.
The super-ratio is only mildly affected by this~\cite{Strauch13PC}, however
the induced polarization can vary substantially.  According to Ref.~\cite{Malace:2011}
the correction is $\lesssim 20\%$, and additional details of how the correction
is made can be found in that work.  In the figures, the curves labeled ``OPT( no CH-EX)''
and ``OPT'' both use one-body electromagnetic currents, the only
difference being that in the ``OPT( no CH-EX)'' calculation the charge-exchange
terms in the optical potential are ignored.  The curves labeled ``OPT+MEC''
include the full optical potential as well as the MEC contributions, while
the curves labeled ``GLB'' correspond to results obtained in the Glauber
approximation with one-body currents.  The statistical errors associated with
the Monte Carlo integrations are only shown for the ``OPT+MEC''
calculation, they are similar in the other cases.  Note that these
errors are smaller than those reported in Ref.~\cite{Schiavilla05b} because
of the larger number of configurations in the present random
walk.
 
The present calculation differs from that reported in Ref.~\cite{Schiavilla05b}
in two respects: i) the spin-orbit term in the optical potential, which is poorly
determined~\cite{Schiavilla05b}, has been constrained here by fitting the
precise induced polarization data obtained in Ref.~\cite{Malace:2011},
and ii) calculations of the super-ratio and induced polarization have
also been carried out in the Glauber approximation (including up to triple
rescattering).  In reference to the calculations based on the optical
potential the discussion and ensuing conclusions are similar to those presented
in the older study~\cite{Schiavilla05b}: i) charge-exchange FSI effects are
important, ii) the predicted quenching of the super-ratio relative to one
comes about because of these effects and because of MEC contributions,
and iii) this quenching is in reasonable agreement with that observed in the
older~\cite{Strauch03} as well as in the more recent and accurate~\cite{Paolone10} data.

The ``GLB'' calculation is at variance with data, particularly at lower $Q^2$. 
While it reproduces the magnitudes of the observables, it has the wrong sign
for $P_y$  and increases the super-ratio relative to one.  However,
we note that for the data in the low $Q^2$ region the proton lab kinetic
energies may be too small for the viability of the Glauber treatment
of FSI, for example at $Q^2=1$ (GeV/c)$^2$ this energy is $\simeq 0.55$
GeV.

\section{Conclusions}
\label{sec:concl}
In this study we have expanded and built upon the work of
Refs.~\cite{Schiavilla05a,Schiavilla05b}, and have calculated
observables for the processes \heedp ~and \heedpfour.  We
have updated the $NN$ amplitudes, which describe FSI within
a Glauber approximation, to include more realistic parameterizations
available from SAID, valid over the entire angular region.  In addition
to the SAID parameterizations we also implemented a minimal $NN$
amplitude, which includes no spin dependence and is only valid
in the forward direction,  allowing for a valuable analysis of the
$NN$ model dependence entering the calculation.  Comparisons
were made to available experimental data, and the theoretical
results are in good agreement with them.

In the case of the \heedp~reaction we have compared several model-dependent
effects which can affect the results significantly.  Among these effects, FSI are of utmost
importance.  Contributions from MEC, while small in some cases, can play a large
role in other observables or kinematical regimes.  We also investigated the importance
of including both the full spin dependence in the profile operator and double rescattering
in the Glauber approximation.  Neglecting either of these effects will have a detrimental
impact on the calculation.

For the \heedpfour~reaction we found that the results obtained
with either the optical potential or Glauber approximation provide
a good description of the data obtained in quasi-parallel kinematics
(PY1 and PY2).  In contrast, the Glauber results overestimate the
data in quasi-perpendicular kinematics (CQ2).  In reference to the
polarization observables measured in the
$^4$He$(\vec{e},e^\prime\vec{p}\,)^3$H reaction, the Glauber
results appear to be severely at variance with data on the
induced polarization $P_y$ and super-ratio
$(P^\prime_x/P^\prime_z)/(P^\prime_x/P^\prime_z)_{\rm PWIA}$,
particularly at low $Q^2$.  In contrast,
these data are reproduced reasonably well in the
calculation based on the optical potential, provided
the latter accounts for charge-exchange FSI effects,
i.e., the coupling between the $p$-$^3$H and $n$-$^3$He
channels.
\section*{Acknowledgments}
We would like to thank R.A.\ Arndt and R.L.\ Workman for
correspondence in regard to the use of the SAID interactive program, 
C. Ciofi degli Atti and H. Morita for the correspondence regarding their $NN$ parameterization, 
and D.\ Higinbotham and B.\ Reitz, and M. Paolone, S.\ Malace and S.\ Strauch, for providing
us with tables of the experimental data for the $^4$He($e,e^\prime p$)$^3$H
and $^4$He($\vec{e},e^\prime \vec{p}\,$)$^3$H reactions, respectively.
This work was supported by the U.S.~Department of Energy, Office of
Nuclear Physics, under contract DE-AC05-06OR23177.  The calculations were
made possible by grants of computing time from the National Energy Research
Scientific Computing Center.
\appendix

\section{$NN$ scattering amplitudes}
\label{app:NNSAIDtoWallace}
This work requires the $NN$ scattering amplitudes as input to describe the FSI. Here we use the $NN$ amplitudes
$\overline{F}^{NN}_{m}(s,t)$ from Eq.~(\ref{eq:fnn}), which are in the Wallace representation~\cite{Wallace81,Wallace83}, 
to produce the profile functions given by Eq.~(\ref{eq:gop}). 
We use a complete set of amplitudes obtained from the SAID analysis and a central (no spin dependence) 
amplitude from Ciofi and Morita 
\cite{ciofihiko_1,ciofihiko_2,ciofihiko_3,ciofihiko_4,ciofihiko_5,ciofihiko_6,ciofihiko_7,ciofihiko_8,ciofihiko_9,ciofihiko_10}. 
Some comments are necessary for each of these choices to clarify their usage in the present work.

It is possible to obtain $NN$ scattering amplitudes in two-dimensional spinor space directly from SAID in the form of the 
Saclay amplitudes which can be easily related to the Wallace form. 
The problem with this is that for lab kinetic energies below 350 MeV these are not in agreement with those obtained from the 
Nijmegen analysis (http://nn-online.org/). 
However, helicity amplitudes can also be obtained directly from SAID and these can then be converted to Saclay amplitudes, 
which are in agreement with the Nijmegen analysis. As a result, we start from the SAID helicity amplitudes. 
These are then converted to the  Fermi invariant amplitudes of Eq.~(\ref{eq:fermi_invariants}) as described in 
Ref.~\cite{JVO_2008_newcalc}. 
The coefficients of the Fermi invariant amplitudes are saved as tables of the five invariant amplitudes and as a function of 
c.m.~angle for laboratory kinetic energies $T_{lab}$ from 0.05 GeV to 1.3 GeV for $pn$ scattering and 0.05 GeV to 3.0 GeV 
for $pp$ scattering. These tables are interpolated using bicubic splines to obtain scattering amplitudes at any 
energy and angle within the tabulated energy range. 
These invariant amplitudes have been used successfully to calculate a number of deuteron electrodisintegration observables 
\cite{JVO_2008_newcalc,JVO_2009_ejec_pol,JVO_2009_tar_pol}. 
For the current work the Fermi invariants are converted to Wallace amplitudes by multiplication by an appropriate matrix. 
Some care has to be used in implementing this approach due to a problem with the production of the helicity amplitudes by SAID. 
In extracting the amplitudes we have specified that at each energy these are given from $\theta_{c.m.}=0^\circ$ to $180^\circ$ 
in steps of $5^\circ$. The resulting amplitudes show a very strong variation at angles near both endpoints resulting in 
differential cross sections that have large spikes near $0^\circ$ and $180^\circ$ that are inconsistent with the scattering data. 
To eliminate this problem, amplitudes at $5^\circ$ and $10^\circ$ are replaced by values obtained from a cubic polynomial 
fixed by data at $0^\circ$, $15^\circ$, $20^\circ$ and $25^\circ$. 
This produces differential cross sections that are in agreement with data. 

\begin{figure}
    \includegraphics[width=16cm]{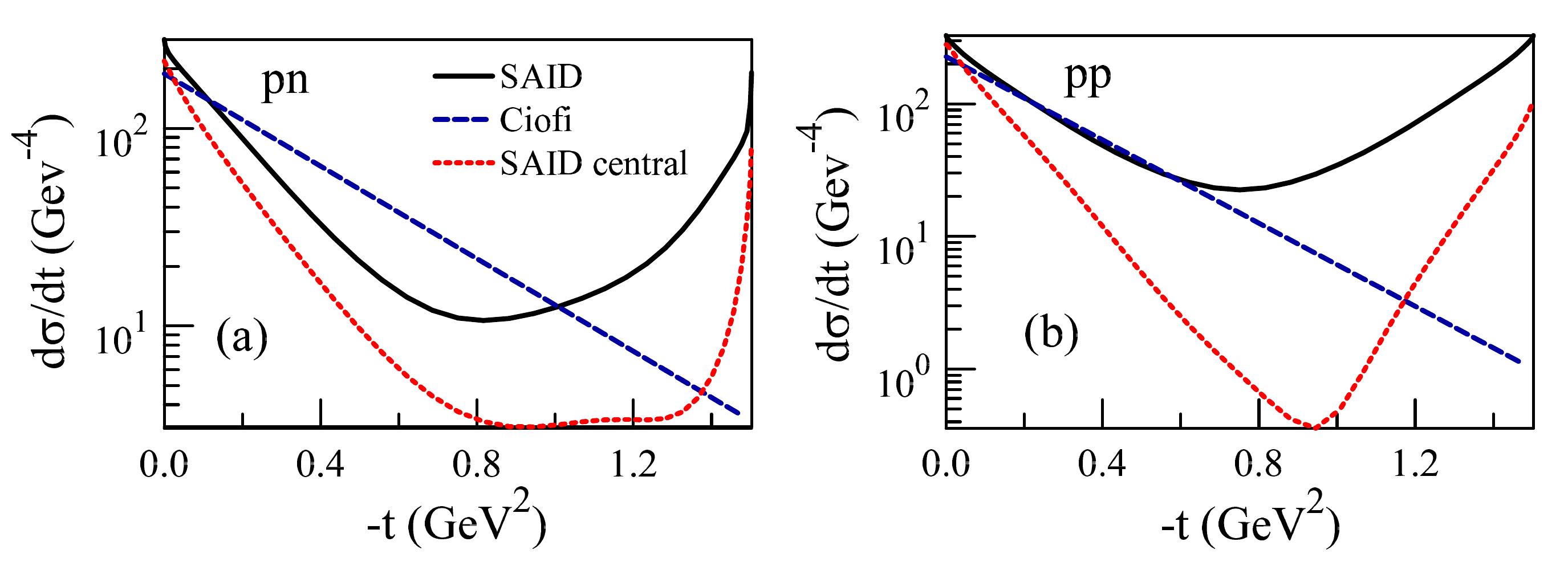}  
    \caption{(color online) Differential cross sections obtained from the SAID (solid line), Ciofi (dashed line) and the central contribution from SAID (short dashed line)  
    amplitudes for (a) $pn$ scattering and (b) $pp$ scattering for $T_{lab}=0.8\ {\rm GeV}$ ($s=5.03\ {\rm GeV^2}$). }
    \label{fig:dsig_dt}
\end{figure}

Ciofi and Morita use only a single spin-independent amplitude of the form
\begin{equation}
\overline{F}^{NN}_1(s,\bar{\bm{k}}^2)=-i\,\sigma_{tot}(s)\left[1-i\alpha(s)\right] {\rm e}^{-\beta\,\bar{\bm{k}}^2}\label{eq:cioffiAmp}
\end{equation} 
where $\sigma_{tot}$ is the total $NN$ cross section, $\alpha$ is a ratio of the real to imaginary part of the 
forward scattering amplitude (often referred to as $\rho$) and $\beta$ is determined by calculating the 
total elastic cross section from Eq.~(\ref{eq:cioffiAmp}) giving
\begin{equation}
\beta=\frac{\sigma_{tot}^2(s)}{32\pi\, \sigma_{el}^2(s)}\left[ 1+\alpha^2(s)\right]\,.\label{eq:beta}
\end{equation}
The quantities $\sigma_{tot}$, $\alpha=\rho$ and $\sigma_{el}$ can be obtained from either the PDG or from SAID. 
Differential cross sections using the SAID and Ciofi amplitudes are shown for $pn$ scattering in Fig. \ref{fig:dsig_dt}(a) 
and for $pp$ scattering in Fig. \ref{fig:dsig_dt}(b) for the full kinematically allowed range in $t=-\bar{\bm{k}}^2$.
Note that while the SAID and Ciofi results are similar in the forward direction for $pp$ scattering, 
this is not the case for $pn$ scattering.
The problem here is in determining $\beta$. 
The total elastic cross section for $pp$ scattering is completely described by integrating from $0^\circ$ to $90^\circ$ 
since for indistinguishable protons each scattering in the c.m.~frame will result in one proton in the forward direction 
and one in the backward direction. 
This is not the case for $pn$ scattering since a forward scattering proton will be associated with a 
backward scattering neutron and a backward scattering proton will be associated with a forward scattering neutron. 
The total elastic $pn$ cross section requires integration from $0^\circ$ to $180^\circ$. 
In the case of Fig. \ref{fig:dsig_dt}(b) the total $pp$ elastic cross section corresponds 
to integrating the differential cross section over half of the range in $t$, 
while for Fig. \ref{fig:dsig_dt}(a) the total $pn$ elastic cross section corresponds to integrating the 
differential cross section over the complete range in $t$. By including the contributions from backward scattering protons, 
the total $pn$ elastic cross section is larger than would be required to fit the data in the forward direction 
resulting in a smaller value of $\beta$ as given by Eq.~(\ref{eq:beta}).
Note also that the values of the cross section calculated from the Ciofi amplitudes are smaller than those obtained from the 
SAID amplitudes at $t=0$ due to the contributions from spin-dependent amplitudes at this point. 

The third calculation shown in Figs. \ref{fig:dsig_dt}(a) and \ref{fig:dsig_dt}(b) shows the contribution of only the central part of the SAID amplitudes to the differential cross section. This clearly shows that the spin-dependent amplitudes provide a significant part of the cross section and that the method used by Ciofi transfers part of this strength into the central amplitude.

Fortran90 modules are available from Ford and Van Orden (FVO) which calculate the invariant functions, ${\cal F}_m^{NN}(s,t)$, 
given in Eq.~(\ref{eq:fermi_invariants}). 
The subroutines can provide the amplitudes for a variety of models depending on the energies desired as well as the 
complexity of the model. At a basic level there is a parametrization available from Ciofi and Morita 
\cite{ciofihiko_1,ciofihiko_2,ciofihiko_3,ciofihiko_4,ciofihiko_5,ciofihiko_6,ciofihiko_7,ciofihiko_8,ciofihiko_9,ciofihiko_10} 
describing the $NN$ system with a single amplitude with no spin dependence. 
Using this amplitude provides a useful comparison for studying how the FSI model dependence, specifically spin dependence, 
contributes to a calculation. 
Next one can choose the Wallace parametrization \cite{Wallace81}, which incorporates spin dependence, 
but is only valid at small angles. This model was utilized in an earlier work \cite{Schiavilla05a} but, due to
the limitation above, is not used in this work. 
There are two parametrizations available which include all spin dependence and are valid over the entire angular region. 
These are the SAID model \cite{Arndt00,Arndt07,SAIDdata} valid for $s<5.4$ (GeV$^2$), 
and a Regge model \cite{FVO_Reggemodel} valid for $s>5.4$ (GeV$^2$). 
In this work we consider all spin dependence of the FSI, and the energies of interest are those of the SAID approach.
For all models the amplitudes are converted first into Fermi invariant functions with a consistent normalization. 
The Fermi invariants from any model can be used directly or converted to helicity amplitudes, Saclay amplitudes or Wallace amplitudes.

As discussed above, for the SAID analysis the five independent helicity amplitudes can be obtained on a tabulated grid for the 
c.m.~energy ($ \overline{E}$) and angle ($\overline{\theta}$). 
For convenience we work with the Mandelstam variables which are related to the $NN$ c.m.~energy and angle by 
\begin{align}
s &= 4\overline{E}^2 \\
t &= -\overline{{\bf k}}^2 = -\frac{s-4\,m^2}{2}[1-\cos(\overline{\theta})].
\end{align}
If the amplitudes are extracted in units of fm there is a normalization relation between the SAID and FVO conventions,
\begin{equation}
T^{FVO}_{\lambda'_1,\lambda'_2;\lambda_1,\lambda_2}(s,t) = -\frac{4\pi\sqrt{s}}{\hbar c \, m^2}
T^{SAID}_{\lambda'_1,\lambda'_2;\lambda_1,\lambda_2}(s,t)
\end{equation}
 The invariants can then be obtained using,
\begin{equation}
\left( \begin{array}{c}
{ \cal F}^{NN}_{S}(s,t)  \\ 
{ \cal F}^{NN}_{V}(s,t) \\ 
{ \cal F}^{NN}_{T}(s,t) \\ 
{ \cal F}^{NN}_{P}(s,t) \\ 
{ \cal F}^{NN}_{A}(s,t)      
\end{array} \right) = \frac{1}{s-4\, m^2} {M^{HtoI}}
\left( \begin{array}{l}
T_{\frac{1}{2},\frac{1}{2};\frac{1}{2},\frac{1}{2}}(s,t)  \\ 
T_{\frac{1}{2},\frac{1}{2};\frac{1}{2},-\frac{1}{2}}(s,t) \\ 
T_{\frac{1}{2},-\frac{1}{2};\frac{1}{2},-\frac{1}{2}}(s,t) \\ 
T_{\frac{1}{2},\frac{1}{2};-\frac{1}{2},-\frac{1}{2}}(s,t) \\ 
T_{\frac{1}{2},-\frac{1}{2};-\frac{1}{2},\frac{1}{2}}(s,t)      
\end{array} \right) ,
\end{equation}
where the matrix $M^{HtoI}$ and additional details of this discussion can be found in the Appendix of
Ref.~\cite{JVO_2008_newcalc}.

Once the invariant functions are obtained we need to represent the amplitudes in the Wallace form so that the 
Glauber profile operator can be calculated. Normalization between the FVO convention and the convention used in this work is given as,
\begin{equation}
T_{\lambda'_1,\lambda'_2;\lambda_1,\lambda_2}(s,t) = \frac{i\hbar c\, m^2}{2 \pi \sqrt{s(s-4\,m^2)}}
T^{FVO}_{\lambda'_1,\lambda'_2;\lambda_1,\lambda_2}(s,t)
\end{equation}
It is straightforward to transform from the invariant functions to the Wallace form via another matrix multiplication,
\begin{equation}
\left( \begin{array}{c}
\overline{F}^{NN}_{1}(s,t)  \\ 
\overline{F}^{NN}_{2}(s,t) \\ 
\overline{F}^{NN}_{3}(s,t) \\ 
\overline{F}^{NN}_{4}(s,t) \\ 
\overline{F}^{NN}_{5}(s,t)      
\end{array} \right) = \frac{i\hbar c \, m^2}{2 \pi \sqrt{s(s-4\, m^2)}} M^{ItoW}
\left( \begin{array}{c}
{ \cal F}^{NN}_{S}(s,t)  \\ 
{ \cal F}^{NN}_{V}(s,t) \\ 
{ \cal F}^{NN}_{T}(s,t) \\ 
{ \cal F}^{NN}_{P}(s,t) \\ 
{ \cal F}^{NN}_{A}(s,t)  
\end{array} \right) ,
\end{equation}
where the matrix $M^{ItoW}$ is given below and was obtained from \cite{Wallace83}. In Appendix \ref{app:flab} 
we show how these amplitudes can be boosted to the rescattering frame (which is in practice taken as the lab
frame, see discussion in Sec.~\ref{sec:glb}), 
and the profile operator can then be calculated from the boosted amplitudes.  
The matrix elements are:
\begin{align} 
M^{ItoW}_{11} &= \frac{(-4m^2 - 2m\sqrt{s} + t)^2}{4m^2(2m + \sqrt{s})^2} \nonumber \\
M^{ItoW}_{12}     &= \frac{(-16m^4 - 16m^3\sqrt{s} + 2s^2 + 3st + t^2 + 4m^2(s + t) + 8m\sqrt{s}(s + t))}{4m^2(2m + \sqrt{s})^2} \nonumber \\
M^{ItoW}_{13} &= \frac{t(4m\sqrt{s} + 2s + t)}{2m^2(2m + \sqrt{s})^2} \nonumber \\ 
M^{ItoW}_{14} &= 0  \nonumber \\
M^{ItoW}_{15} &= \frac{(4m^2 - s - t)t}{4m^2(2m + \sqrt{s})^2} \nonumber \\
M^{ItoW}_{21} &= \frac{t(-4m^2 + s + t)}{4m^2(2m + \sqrt{s})^2}  \nonumber \\
M^{ItoW}_{22} &= \frac{t(4m\sqrt{s} + 2s + t)}{4m^2(2m + \sqrt{s})^2} \nonumber \\
M^{ItoW}_{23} &= \frac{-16m^4 - 16m^3\sqrt{s} + 2s^2 + 3st + t^2 + 4m^2(s + t) 
       		+ 8m\sqrt{s}(s + t)}{2m^2(2m + \sqrt{s})^2}  \nonumber \\
M^{ItoW}_{24} &= 0 \nonumber \\
M^{ItoW}_{25} &= -\frac{(-4m^2 - 2m\sqrt{s} + t)^2}{4m^2(2m + \sqrt{s})^2}   \nonumber \\
M^{ItoW}_{31} &= \frac{(4m^2 + 2m\sqrt{s} - t)\sqrt{-4m^2 + s + t}}{4m^2(2m + \sqrt{s})^2} \nonumber \\
M^{ItoW}_{32} &=  -\frac{(4m^2 + 6m\sqrt{s} + 2s + t)\sqrt{-4m^2 + s + t}}{4m^2(2m + \sqrt{s})^2} \nonumber \\
M^{ItoW}_{33} &= - \frac{(4m^2 + 6m\sqrt{s} + 2s +t)\sqrt{-4m^2 + s + t}}{2m^2(2m + \sqrt{s})^2} \nonumber \\
M^{ItoW}_{34} &= 0 \nonumber \\
M^{ItoW}_{35} &= \frac{(-4m^2 - 2m\sqrt{s} + t)\sqrt{-4m^2 + s + t}}{4m^2(2m + \sqrt{s})^2}  \nonumber \\
M^{ItoW}_{41} &= \frac{-4m^2 + s + t}{4m^2(2m + \sqrt{s})^2} \nonumber \\
M^{ItoW}_{42} &= \frac{4m\sqrt{s} + 2s + t}{4m^2(2m + \sqrt{s})^2}  \nonumber \\
M^{ItoW}_{43} &= \frac{4m\sqrt{s} + 2s + t}{2m^2(2m + \sqrt{s})^2} \nonumber \\
M^{ItoW}_{44} &= - \frac{1}{4m^2} \nonumber \\
M^{ItoW}_{45} &= \frac{8m^2 + 4m\sqrt{s} - t}{4m^2(2m + \sqrt{s})^2} \nonumber \\
M^{ItoW}_{51} &= \frac{(4m^2 - s - t)t}{4m^2(2m + \sqrt{s})^2} \nonumber \\
M^{ItoW}_{52} &= \frac{(4m^2 - s - t)t}{4m^2(2m + \sqrt{s})^2} \nonumber \\
M^{ItoW}_{53} &= -\frac{-32m^4 - 32m^3\sqrt{s} +  2s^2 + 4m^2t + 3st + t^2 + 8m\sqrt{s}(s + t)}{2m^2(2m + \sqrt{s})^2}  \nonumber \\
M^{ItoW}_{54} &= 0 \nonumber \\
M^{ItoW}_{55} &= \frac{32m^4 + 32m^3\sqrt{s} - 2s^2 - 12m^2t - st + t^2 - 8m\sqrt{s}(s + t)}{4m^2(2m + \sqrt{s})^2}  \ .
\end{align}
\section{From the c.m.~to the lab frame}
\label{app:flab}

The elastic scattering amplitude in the lab frame is written as
\begin{equation}
(2 {\rm i}\, p)^{-1}\, F^{NN}_{ij}({\bf k},s)
=\sum_{m=1}^8 F^{NN}_m(s,{\bf k}^{\, 2}) O^m_{ij} \ ,
\label{eq:flab}
\end{equation}
where the eight operators $O^m_{ij}$ are taken as
\begin{equation}
O^{\,m=1,\dots,8}_{ij}=1\, ,\,  {\bm \sigma}_i \cdot {\bm \sigma}_j\, ,\,
{\rm i}\, {\bm \sigma}_i \cdot {\bf k} \times \hat {\bf p}\, ,\,
{\rm i}\, {\bm \sigma}_j \cdot {\bf k} \times \hat {\bf p}\, ,\,
{\bm \sigma}_i \cdot {\bf k} \, {\bm \sigma}_j \cdot {\bf k}\, ,\,
{\bm \sigma}_i \cdot \hat{\bf p} \, {\bm \sigma}_j \cdot \hat{\bf p} \, , \,
i\, {\bm \sigma}_i \cdot {\bf k} \, {\bm \sigma}_j \cdot \hat{\bf p} \, , \,
i\, {\bm \sigma}_i \cdot \hat{\bf p} \, {\bm \sigma}_j \cdot {\bf k} \ .
\label{eq:oplab}
\end{equation}
Here ${\bf p}$ is the momentum of the initial fast nucleon and in
the eikonal limit the momentum transfer ${\bf k}$ is perpendicular 
to ${\bf p}$.  The functions $F^{NN}_{m=1,\dots, 8}$ are then obtained
as linear combinations of the invariant functions
${\cal F}_{m=1,\dots, 5}^{NN}$,
\begin{equation}
F^{NN}_m=\sum_{n=1}^5 L_{mn}\, {\cal F}^{NN}_n \ ,
\end{equation}
where the 8$\times$5 matrix $L$ is given by 
\vskip 0.7cm

\noindent
\centerline{
$
L= \left[ \begin{array}{ccccc}
1-\frac{p^2}{w_p w_{p-k}} & 1+\frac{p^2}{w_p w_{p-k}}+\frac{k^2}{w_{p-k} w_k}&
-\frac{2 k^2}{w_{p-k} w_k} & 0  & 0  \\
0    & -\frac{k^2}{w_{p-k} w_k} & 2\left(1+\frac{p^2}{w_p w_{p-k}}+
\frac{k^2}{w_{p-k} w_k}\right)& 0 & -1+\frac{p^2}{w_p w_{p-k}} \\
\frac{p}{w_p w_{p-k}}  & -\frac{p}{w_p w_{p-k}}\left(1+\frac{w_-}{w_k}\right) &
-\frac{2p}{w_p w_{p-k}} \frac{w_+}{w_k}    &  0 & 0  \\
0 &-\frac{p}{w_p w_{p-k}} \frac{w_+}{w_k} &
-\frac{2p}{w_p w_{p-k}}\left(1+\frac{w_-}{w_k}\right) & 0 &
-\frac{p}{w_p w_{p-k}} \\
0 & \frac{1}{w_{p-k} w_k} & -\frac{2}{w_{p-k} w_k} & -\frac{1}{w_{p-k} w_k} &
-\frac{1}{w_{p-k} w_k} \\
0 & 0 & -\frac{4p^2}{w_p w_{p-k}} & 0 & -\frac{2p^2}{w_p w_{p-k}} \\
0 & -i\frac{p}{w_p w_{p-k}}\frac{w_-}{w_k} &
-i\frac{2p}{w_p w_{p-k}}\left( 1+\frac{w_+}{w_k}\right) & 0 &
-i\frac{p}{w_p w_{p-k}} \\
0 & 0 & -i\frac{2p}{w_p w_{p-k}} & -i\frac{p}{w_p w_{p-k}}\frac{w_-}{w_k} &
-i\frac{p}{w_p w_{p-k}}\left( 1+\frac{w_+}{w_k}\right)
\end{array} \right] \ ,
$ }
\vskip 0.7cm
\noindent and the factors $E_q$ and $w_q$ are defined as
$E_q\equiv \sqrt{{\bf q}^2+m^2}$ and $w_q\equiv E_q+m$,
with ${\bf q}={\bf p}$, ${\bf k}$, ${\bf p}-{\bf k}$,
and $w_{\pm}\equiv w_{p-k}\pm w_p$,

The $N$$N$ profile operator $\Gamma^{NN}_{ij}$ is obtained from
Eq.~(\ref{eq:gop}) by replacing $\Gamma^{(m)}_{ij}$ with
$\Gamma^{(m)}_{NN}$ for $m=1,\dots, 8$.  The functions
$\Gamma^{(m)}_{NN}$ are in turn derived from Bessel transforms
of the $F^{NN}_{m}$ amplitudes.  We find:
\begin{equation}
\Gamma^{(m)}_{NN}(b;s)=2\, p^2\int_{-1}^1 {\rm d}x\, J_0(kb)\,
F^{NN}_{m}(k^2;s)
\end{equation}
for $m=1,6$;
\begin{equation}
\Gamma^{(m)}_{NN}(b;s)=\frac{ 2\, p^2}{b}
\int_{-1}^1 {\rm d}x\, k\, J_1(kb)\, F^{NN}_{m}(k^2;s) 
\end{equation}
for $m=3,4,7,8$; and lastly
\begin{eqnarray}
\Gamma^{(2)}_{NN}(b;s)&=&2\, p^2\int_{-1}^1 {\rm d}x\,
J_0(kb)\, F^{NN}_{2}(k^2;s) 
+\frac{ 2\, p^2}{b} \int_{-1}^1 {\rm d}x\, k\,
J_1(kb)\, F^{NN}_{5}(k^2;s) \ , \\
\Gamma^{(5)}_{NN}(b;s)&=&\frac{ 2\, p^2}{b^2}
\int_{-1}^1 {\rm d}x\, k^2\, \left[ J_0(kb)-
\frac{2}{k\, b}J_1(kb)\right] \, F^{NN}_{5}(k^2;s) \ .
\end{eqnarray}
In obtaining the integrals above, we made the variable change
$k \rightarrow 2\,p\, {\rm sin}(\theta/2)=
2\, p\, \sqrt{(1-x)/2}$ with $x={\rm cos}\,\theta$.
\bibliography{heedp}

\end{document}